%% file: main.tex
\documentclass[final,3p,times]{elsarticle}
\usepackage{tabularx}
\usepackage{amsmath}
\usepackage{amsfonts}
\usepackage{amsthm}
\usepackage{graphicx}
\usepackage{subcaption}
\usepackage{amssymb}
\usepackage{color}
\usepackage{bm}
\usepackage{hyperref}
\hypersetup{
colorlinks   = true,
citecolor    = blue,
linkcolor    = blue}

\DeclareSymbolFont{CMMI}{OML}{ccm}{m}{it}
\DeclareMathSymbol{v}{\mathalpha}{CMMI}{"76}

\input{commandsHeader.tex}

\journal{Computational Physics Communications}

\begin{document}

\begin{frontmatter}

\title{Conservative discontinuous Galerkin scheme of a gyro-averaged Dougherty collision operator}
\author[mit]{M. Francisquez}
\author[ga]{T.~N. Bernard}
\author[PU]{N.~R. Mandell}
\author[pppl]{G.~W. Hammett}
\author[pppl]{A. Hakim}
\address[mit]{MIT Plasma Science and Fusion Center, Cambridge, MA, 02139}
\address[ga]{General Atomics, PO Box 85608, San Diego, CA 92186}
\address[PU]{Department of Astrophysical Sciences, Princeton University, Princeton, NJ 08544}
\address[pppl]{Princeton Plasma Physics Laboratory, Princeton, NJ 08543}

\begin{abstract}
A conservative discontinuous Galerkin scheme for a nonlinear Dougherty collision operator in full-$f$ long-wavelength gyrokinetics is presented. Analytically this model operator has the advective-diffusive form of Fokker-Planck operators, it has a non-decreasing entropy functional, and conserves particles, momentum and energy. Discretely these conservative properties are maintained exactly as well, independent of numerical resolution. In this work the phase space discretization is performed using a novel version of the discontinuous Galerkin scheme, carefully constructed using concepts of weak equality and recovery. Discrete time advancement is carried out with an explicit time-stepping algorithm, whose stability limits we explore. The formulation and implementation within the long-wavelength gyrokinetic solver of \gkeyll~are validated with relaxation tests, collisional Landau-damping benchmarks and the study of 5D gyrokinetic turbulence on helical, open field lines.
\end{abstract}

\end{frontmatter}
%\tableofcontents

\input{introduction}
\input{theory}
\input{scheme}
\input{stability}
\input{results}
\input{conclusions}

\section*{Acknowledgements} 
We thank Darin Ernst and James Juno for useful discussions on collision operators and DG, and Petr Cagas for the development of the \texttt{postgkyl}~data visualization tool. The simulations presented here were carried out at the Texas Advanced Computing Center, the Dartmouth Discovery cluster, and MIT's Engaging cluster, so we wish to thank the support teams at these facilities for their work in maintaining these systems. MF is supported by DOE contract DE-FC02-08ER54966. TNB was supported by DOE contract DE-FG02-04ER-54742, through the Institute of Fusion Studies at the University of Texas at Austin, and is currently supported by DOE contract DE-FG02-95ER54309. NRM is supported by the DOE CSGF program, provided under grant DE-FG02-97ER25308. AH and GWH are supported by the High-Fidelity Boundary Plasma Simulation SciDAC Project, part of the DOE Scientific Discovery Through Advanced Computing (SciDAC) program, through DOE contract DE-AC02-09CH11466 for the Princeton Plasma Physics Laboratory. AH is also supported by the Air Force Office of Scientific Research under contract FA9550-15-1-0193.

\appendix
\input{appendixA.tex}

\section*{References}
\bibliographystyle{elsarticle-num} 
\bibliography{main.bib}

\end{document}

%% file: commandsHeader.tex
\renewcommand{\v}[1]{\ensuremath{\mbox{\boldmath$ #1 $}}} % for vectors
\newcommand{\gv}[1]{\ensuremath{\mbox{\boldmath$ #1 $}}} 
\newcommand{\uv}[1]{\ensuremath{\mathbf{\hat{#1}}}} % for unit vector vector
\newcommand{\avg}[1]{\left< #1 \right>} % for average
\renewcommand{\d}[2]{\frac{d #1}{d #2}} % for derivatives
\newcommand{\pd}[2]{\frac{\partial #1}{\partial #2}} 
\newcommand{\pdd}[2]{\frac{\partial^2 #1}{\partial #2^2}} 
\newcommand{\pddo}[3]{\frac{\partial^2 #1}{\partial #2 \partial #3}} 
\newcommand{\grad}[1]{\nabla #1} % for gradient
\newcommand{\gradperp}[1]{\nabla_\perp #1} % for perpendicular gradient
\renewcommand{\div}[1]{\nabla \cdot #1} % for divergence
\newcommand{\curl}[1]{\nabla \times #1} % for curl
\newcommand{\inProd}[2]{{\left\langle #1, #2 \right\rangle}} % inner product.
\newcommand{\kron}[2]{\delta_{#1,#2}} % Kronecker delta.
\newcommand{\gkcoll}[1]{\mathcal{C}[#1]} % Collision operator.
\newcommand{\coll}[1]{\left(\pd{ #1}{t}\right)_c} % Collision operator.
\newcommand{\zhat}{\uv{z}}
\newcommand{\txtS}[2]{#1_{\mathrm{#2}}}

\newcommand{\pfrac}[2]{\frac{\partial #1}{\partial #2}}

\newcommand{\hermit}[1]{\mathrm{H}_{#1}}

\newcommand{\omegaci}{\Omega_{i}}

\newcommand{\bhat}{\gv{b}}
\newcommand{\kpar}{k_\parallel}
\newcommand{\kperp}{k_{\perp}}
\newcommand{\nuee}{\nu_{ee}}

\newcommand{\nuie}{\nu_{ie}}
\newcommand{\nuii}{\nu_{ii}}
\newcommand{\uparZ}{u_{\parallel,0}}
\newcommand{\uparO}{u_{\parallel,1}}
\newcommand{\vtZ}{v_{t,0}}
\newcommand{\vtO}{v_{t,1}}
\newcommand{\rhoi}{\rho_i}
\newcommand{\rhos}{\rho_s}

\newcommand{\dx}{\mathrm{d}x}
\newcommand{\Dx}{\Delta x}
\newcommand{\Dz}{\Delta z}
\newcommand{\Dv}{\Delta v}

\newcommand{\Dxcell}{\txtS{\Delta x}{cell}}
\newcommand{\vt}{v_{t}}
\newcommand{\vts}{v_{ts}}
\newcommand{\vte}{v_{te}}
\newcommand{\vti}{v_{ti}}

\DeclareSymbolFont{CMBI}{OML}{ccm}{b}{it}
\DeclareMathSymbol{\vv}{\mathalpha}{CMBI}{"76}
\newcommand{\vi}{v_i}
\newcommand{\vj}{v_j}

\newcommand{\divv}[1]{\nabla_{\vv} \cdot #1}
\newcommand{\vE}{\v{E}}
\newcommand{\vB}{\v{B}}

\newcommand{\intInf}{\int_{-\infty}^{\infty}}

\newcommand{\dThvv}{\mathrm{d}^3\vv}

\newcommand{\Dt}{{\Delta}t}
\newcommand{\xmax}{x_{\mathrm{max}}}
\newcommand{\xmin}{x_{\mathrm{min}}}

\newcommand{\vcen}{\check{v}}
\newcommand{\eig}{\lambda}
\newcommand{\eigmax}{\txtS{\lambda}{max}}
\newcommand{\Lx}{L_x}

\newcommand{\adv}{\mathrm{adv}}
\newcommand{\dif}{\mathrm{dif}}

\newcommand{\cfl}{\ensuremath{\mathrm{CFL}}} % CFL factor in determination of dt.
\newcommand{\Np}{N_p} % Number of phase space monomials.
\newcommand{\frec}{\hat{f}}	% Recovery polynomial.
\newcommand{\Jac}{\mathcal{J}}

\newcommand{\vpar}{v_\parallel}
\newcommand{\vperp}{v_\perp}
\newcommand{\upar}{u_\parallel}
\newcommand{\vparmax}{v_{\parallel,\max}}
\newcommand{\vparmin}{v_{\parallel,\min}}
\newcommand{\mumax}{\mu_{\max}}
\newcommand{\mumin}{\mu_{\min}}

\newcommand{\Dvpar}{\Delta v_\parallel}
\newcommand{\Dmu}{\Delta\mu}

\newcommand{\dvpar}{\mathrm{d}v_\parallel}
\newcommand{\dmu}{\mathrm{d}\mu}
\newcommand{\mmax}{\txtS{m}{max}}
\newcommand{\dxdmu}{\thinspace\mathrm{d}x\thinspace\mathrm{d}\mu}
\newcommand{\dxdv}{\thinspace\mathrm{d}x\thinspace\mathrm{d}\vpar}
\newcommand{\dxdvdmu}{\thinspace\mathrm{d}x\thinspace\mathrm{d}\vpar\thinspace\mathrm{d}\mu}
\newcommand{\dvdmu}{\thinspace\mathrm{d}\vpar\thinspace\mathrm{d}\mu}

\newcommand{\eqr}[1]{(#1)}

\newcommand{\eqsr}[2]{(#1)-(#2)}
\newcommand{\Eqsr}[2]{(#1)-(#2)}
\newcommand{\figr}[1]{figure\thinspace(#1)}
\newcommand{\figra}[2]{figure\thinspace(#1#2)}

\newcommand{\Figr}[1]{Figure\thinspace(#1)}
\newcommand{\Figra}[2]{Figure\thinspace(#1#2)}

\newcommand{\gkeyll}{{\textsc{Gkeyll}}}

\newcommand{\mFPO}{{GkLBO}}
\newcommand{\xgc}{{\tt XGC}}

%% file: introduction.tex
\section{Introduction} \label{sec:intro}

Many phenomena in plasma physics require kinetic treatment, meaning one must solve for the time-evolution of the particle distribution function in position-velocity phase space. For example, in collisionless astrophysical shocks the electrons and ions do not thermalize with each other nor with themselves via collisions on the time-scales of interest. Such systems are best modeled by the Vlasov-Fokker-Planck (Vlasov-FPO) equation for the particle probability distribution $f_s(t,\v{x},\vv)$:
\begin{equation} \label{eq:boltzmann}
\pd{f_s}{t}+\div{\left(\vv f_s\right)}+\divv{\left(\v{a}_s f_s\right)} = \coll{f_s},
\end{equation}
where $\v{a}_s=\left(q_s/m_s\right)\left(\vE+\vv\times\vB\right)$ is the acceleration due to the Lorentz force, and $q_s$ and $m_s$ the charge and mass of species $s$, respectively. One must simultaneously solve Maxwell's equations to obtain the fields $\vE$ and $\vB$. For most plasmas of interest, the cumulative effect of frequent small-angle collisions dominates over that of rare ballistic, ``large-angle'' collisions, which are more common in neutral gases or fluids. Then the effect of binary encounters can be modeled by the Fokker-Planck operator (FPO), which in Rosenbluth form~\cite{Rosenbluth1957} is
\begin{equation}\label{eq:FProsenbluth}
\coll{f_s} = -\pd{}{\vi}\avg{\Delta\vi}_sf_s +\frac{1}{2}\pddo{}{\vi}{\vj}\avg{\Delta\vi\Delta\vj}_s f_s.
\end{equation}
Here $\avg{\Delta\vi}_s$ indicates the average increments per unit time of the $i$-th component of the velocity of species $s$. Computation of such increments involves integrals of the distribution functions, indicating that \eqr{\ref{eq:FProsenbluth}} is a nonlinear integro-differential operator.

\Eqsr{\ref{eq:boltzmann}}{\ref{eq:FProsenbluth}} and the accompanying Maxwell equations pose a formidable six-dimensional problem spanning a wide range of spatial and temporal scales, which is often numerically intractable. For magnetized environments and phenomena that evolve on a timescale much slower than the rapid particle gyration around the magnetic field, one can use a gyrokinetic reduction to five dimensions by systematically averaging over the fast gyromotion~\cite{sugama2000,brizard2007,krommes_2012}. Additionally, in direct numerical simulations, the gyrokinetic equation permits the use of a larger time-step and coarser grids than the six-dimensional Vlasov-FPO equation. This formulation results in a change of variables from the phase-space variables ($\v{x},\v{v}$) to the gyro-averaged particle position, or gyrocenter, phase-space variables $(\v{R},\vpar,\mu)$, where $\v{R}$ is the gyrocenter coordinate, $\vpar$ is the velocity component parallel to the background magnetic field, and $\mu=m_s\vperp^2/2B$ is the magnetic moment. Here we focus on the electrostatic gyrokinetic model evolving the gyrocenter distribution function $f_s(t,\v{R},\vpar,\mu)$, though the collision operator formulation and implementation presented here can also be incorporated in electromagnetic gyrokinetics~\cite{Mandell2019}. The gyrokinetic equation in this case refers to
\begin{equation} \label{eq:gk-boltzmann}
\pd{\Jac f_s}{t}+\div{(\Jac f_s \dot{\v{R}})}+\frac{\partial}{\partial \vpar}( \Jac f_s \dot{\vpar}) = \Jac\gkcoll{f_s},
\end{equation}
where $\Jac=B_\parallel^*$ is the gyrokinetic phase-space Jacobian and $B_\parallel^*=\bhat\cdot \v{B}^*$ is the parallel component of the effective magnetic field $\v{B}^*=\vB+(m_s \vpar/q_s)\curl{\bhat}$, where $\vB = B\thinspace\bhat$ is the background magnetic field. We will approximate $\bhat\cdot\nabla\times\bhat\approx 0$ so that $B_\parallel^*\approx B$. The gyrokinetic nonlinear FPO is represented by $\gkcoll{f_s}$. The phase-space advection velocities $\dot{\v{R}} = \{\v{R},H\}$ and $\dot{v}_\parallel = \{\vpar,H\}$ are defined in terms of the Poisson bracket
\begin{equation}
\{F,G\} = \frac{\vB^*}{m_s B_\parallel^*} \cdot \left( \grad{F}\pd{F}{\vpar} - \pd{F}{\vpar}\grad{G} \right) - \frac{1}{q_s B_\parallel^*} \bhat \cdot \grad{F} \times \grad{G},
\label{eq:gkcoll-pb-eqn}
\end{equation}
where the Hamiltonian
\begin{equation} \label{eq:hamiltonian}
    H_s = \frac{1}{2}m_s v_\parallel^2 + \mu B + q_s \phi
\end{equation}
depends on the electrostatic potential $\phi$. To complete the electrostatic gyrokinetic system, one may solve for the electrostatic potential using the gyroaveraged Poisson equation
\begin{equation} \label{eq:5-poisson}
-\div{\left(\frac{n_{i}^g q_i^2 \rho_{s}^2}{T_{e}}\gradperp{\phi} \right)} = \sigma_g = q_i n_i^g(\v{R},t) - e n_e(\v{R},t),
\end{equation}
where $n_{i}^g$ is the ion gyro-center density, $T_{e}$ is the electron temperature and $\rho_{s}=c_s/\omegaci$ is the ion sound gyro-radius, with $c_s$ being the sound speed and $\omegaci$ the ion gyro-frequency. In this work we limit ourselves to the long-wavelength limit of gyrokinetics, also known as drift-kinetics. Therefore, the true potential $\phi$ appears in \eqr{\ref{eq:hamiltonian}} instead of the gyroaveraged quantity $\avg{\phi}_\alpha$, we use a simple form of the Poisson equation, and we neglect finite Larmor radius (FLR) effects other than the first-order polarization charge density on the left hand side of \eqr{\ref{eq:5-poisson}}. The Poisson equation is also solved using a spatially constant polarization coefficient.

Several non-turbulence codes have used the full nonlinear FPO~\cite{McCoy1981,Killeen1986,Harvey1992,Dorf2014}, and the \xgc~particle-in-cell code is one of the few turbulence modeling applications employing such an operator~\cite{Hager2016}. The nonlinear FPO is less frequently found among turbulence solvers, and in continuum gyrokinetic turbulence many of the collision operators have been linear. These linear operators were developed for $\delta f$ studies~\cite{Catto1977,Abel2008,Catto2009} and preserved key properties of the FPO, like conservation (of particles, momentum and energy) and non-decreasing entropy. More recently, improved collision models for gyrokinetics have been formulated and implemented in several codes~\cite{Sugama2009,Esteve2015,Sugama2019}. They retained important physics, such as velocity-dependent collisionalities and FLR effects, but are still linearized operators.
% and, in most cases, unable to show entropy is a non-decreasing function. 
% gwh: Sugama is much more positive about the
% entropy properties of his operator, so I
% commented out the last part of the above sentence.
With the exception of a few reports~\cite{Esteve2015,Donnel2019}, there are few simplified collision models in full-$f$ gyrokinetic turbulence modeling. Although the exact linearized gyrokinetic FPO has been formulated~\cite{Li2011,Pan2019}, and implemented~\cite{Pan2020}, and the nonlinear one formulated~\cite{Jorge2019}, their cost can be prohibitive for some applications. There is thus a strong interest in developing simple models that capture some of the important physics, yet can be efficiently implemented in numerical simulation.

We present here the discontinous Galerkin (DG) implementation of a nonlinear full-$f$ continuum gyroaveraged model collision operator. We build upon the formulation and implementation of similar algorithms for a Vlasov-Maxwell model operator presented in~\cite{Hakim2019} (overview of DG schemes and background on various FPO solvers also appears therein), and employ the same concepts of weak equivalence, boundary correction and recovery. These algorithms are implemented in the DG gyrokinetic model within the \gkeyll~computational plasma physics framework~\cite{shi2017gyrokinetic,Hakim2019a}; download instructions as well as directions for running input files associated with this work can be found in~\ref{sec:appendixGkeyll}. In section~\ref{sec:theory}, we present the Dougherty collision operator and its properties in continuous, infinite-velocity space. These features motivate the formulation of the discrete operator and its implementation, presented in section~\ref{sec:discrete}. In section~\ref{sec:stability}, we discuss a suitable time-stepping algorithm and some of the challenges associated with the explicit finite-time integration of the collisional gyrokinetic equation. Finally, section~\ref{sec:results} presents a number of relaxation tests showing the accuracy of the time evolution and the steady state solution with this operator, as well as a collisional Landau damping benchmark, and a complex, five-dimensional gyrokinetic simulation of open helical field-line turbulence. Additional discussion and concluding remarks are found in section~\ref{sec:conclusion}.

%% file: theory.tex
\section{The continuous gyroaveraged Dougherty operator} \label{sec:theory}

The full FPO in~\ref{eq:FProsenbluth} can be simplified considerably while keeping its advective-diffusive structure. One choice is to take $\avg{\Delta\vi}_s = -\nu_{ss}\left(\vi-u_{s,i}\right)$, that is, the frictional velocity change of a particle colliding with frequency $\nu_{ss}$ in a fluid of mean velocity $u_{s,i}$. The operator is simplified further by assuming that the Debye length is much smaller than the Larmor radius of a thermal particle, making perturbations within a Debye sphere isotropic. In this case one can show $\avg{\Delta\vi\Delta\vj}_s\to2\nu_{ss}\vts^2\delta_{ij}$, where $\vts^2=T_s/m_s$~\cite{Dougherty1964}. In this work we limit ourselves to collisions between particles of the same species, and we make the additional simplification that the collision frequency is velocity-independent. In reality $\nu$ should decrease as $v^{-3}$ such that the high energy tail of the distribution is increasingly collisionless. Multi-species collisions and velocity-dependent $\nu_{ss}$ will be described in subsequent work.

For the long-wavelength gyrokinetic model in \eqsr{\ref{eq:gk-boltzmann}}{\ref{eq:5-poisson}}, the collision operator is further simplified since $\v{u}_s = u_{\parallel s}\bhat$. Thus, the gyroaveraged Dougherty operator (\mFPO) becomes
\begin{equation} \label{eq:GkFP}
\Jac\gkcoll{f_s} \equiv \nu_{ss}\left\lbrace\pd{}{\vpar}\left[(\vpar - u_{\parallel s}) \Jac f_s + \vts^2 \pd{\Jac f_s}{\vpar}\right] + \pd{}{\mu}\left[2 \mu \Jac f_s + 2\frac{m_s \vts^2}{B} \mu \pd{\Jac f_s}{\mu}\right]\right\rbrace.
\end{equation}
The Jacobian $\Jac=B$ only varies in configuration-space and can be incorporated into the collision operator. The operator in equation~\eqr{\ref{eq:GkFP}} is frequently called the Lenard-Bernstein operator, and since we intend to use it with the (long-wavelength) gyrokinetic solver we refer to it as the \mFPO. Due to its advective-diffusive similarity to the FPO, we intend to employ the concepts and algorithms laid forth here for the full FPO, but in that case the velocity increments would be determined from the Rosenbluth potentials. For simplicity the species subscript `$s$' will henceforth be assumed. The ``primitive" moments $\upar$ and $\vt$ are calculated in terms of the moments of the distribution function $f(t,\v{R},\vpar,\mu)$:
\begin{align}
    M_0 &= \intInf \Jac f(t,\v{R},\vpar,\mu) \thinspace \dThvv, \label{eq:M0} \\
    M_1 &= \intInf \vpar\thinspace \Jac f(t,\v{R},\vpar,\mu) \thinspace \dThvv, \label{eq:M1} \\
    M_2 &= \intInf \left(\vpar^2 + 2\mu B/m\right) \thinspace \Jac f(t,\v{R},\vpar,\mu) \thinspace \dThvv \label{eq:M2},
\end{align}
where $\int^\infty_{-\infty} \dThvv = (2 \pi/m) \int^\infty_{-\infty} \dvpar \int_0^\infty \dmu$ since we are using gyrokinetic coordinates. With the first three moments of the distribution function, the primitive moments (mean velocity and thermal speed) are then calculated using the relations
\begin{align}
\upar M_{0} &= M_{1}, \label{eq:uparEq} \\
\upar M_{1} + 3 \vt^2 M_{0} &= M_{2}. \label{eq:vtSqEq}
\end{align}

The gyrokinetic model as a whole conserves particle number, total momentum and total energy. To show this one must integrate the gyrokinetic equation~\ref{eq:gk-boltzmann} over all phase space. But the continuous \mFPO~analytically conserves particle number, momentum and energy of each species, and to demonstrate such properties it sufficies to take velocity moments of the collision operator alone. Particle number conservation of the \mFPO~is evident in its (velocity) divergence form and assuming that the argument of the divergence vanishes at infinity. Momentum conservation,
\begin{equation}
    \pd{}{t} \intInf m \vpar\thinspace \Jac \gkcoll{f} \thinspace \dThvv = 0,
\end{equation}
can be satisfied as long as the definition
\begin{equation}
-  \intInf m \nu\thinspace (\vpar - \upar) \Jac f \thinspace \dThvv = 0. \label{eq:mom-ct}
\end{equation}
is obeyed, which leads to~\ref{eq:uparEq}. Similarly, conservation of total particle energy
\begin{equation}
    \pd{}{t}\intInf\left(m\vpar^2/2 + \mu B\right)\thinspace\Jac\gkcoll{f} \thinspace \dThvv = 0 
\end{equation}
will also be preserved as long as one satisfies the following relation
\begin{equation}
- \intInf \nu \left[ m \vpar (\vpar - \upar) + 2\mu B - 3 m \vt^2 \right] \Jac f
\thinspace \dThvv
= 0, \label{eq:en-ct}
\end{equation}
and this is equivalent to~\ref{eq:vtSqEq}. Arriving at the above properties and conditions requires that drag terms be integrated by parts once and the diffusion terms twice. One must also use the fact that $f(t,\v{R},\vpar,\mu)\rightarrow 0$ faster than any power of the velocity as $v_\parallel, \mu \rightarrow \infty$. Although these features, and their proofs, are discussed in various other texts, we summarize them to illustrate the nature of the constraints on the discrete \mFPO~in order to arrive at a consertive scheme.

%An analogous discrete operator must conserve these quantities in a discrete sense, accounting for boundary conditions if finite velocity-space extents are used.

Another important feature of a good collision operator we would like to carry over discretely is ensuring entropy is a non-decreasing function and that the system relaxes to a maximum-entropy solution, i.e.~a Maxwellian function. Defining the entropy as $S = -\int_{-\infty}^{\infty} f \ln{f} \thinspace \dThvv$, this means
\begin{equation} \label{eq:entrEv}
\pd{S}{t} = -\int_{-\infty}^{\infty} \pd{f}{t}(\ln f + 1) \thinspace \dThvv \ge 0.
\end{equation}
One can show that the \mFPO~obeys such relation by writing the operator as
\begin{equation}
\Jac\gkcoll{f} = \pd{F_{\vpar}}{\vpar} + \pd{F_\mu}{\mu}
\end{equation}
where 
\begin{align}
F_{\vpar} &= \nu(\vpar - \upar) \Jac f + \nu\vt^2 \pd{\Jac f}{\vpar} \label{eq:gkfpo-para} \\
F_\mu &= 2\nu \mu \Jac f + \nu\frac{2m \vt^2}{B} \mu \pd{\Jac f}{\mu} \label{eq:gkfpo-mu}
\end{align}
Substitute these definitions into \eqr{\ref{eq:entrEv}} and integrate by parts to get
\begin{equation} \label{eq:entrEvNew}
\pd{S}{t} = \int_{-\infty}^{\infty} \frac{1}{\Jac f} \left( F_{\vpar} \pd{f}{\vpar} + F_\mu \pd{f}{\mu} \right) \thinspace d^3\vv,
\end{equation}
assuming that $F_{\vpar} \rightarrow 0$ as $\vpar \rightarrow \pm \infty$ and $F_{\mu} \rightarrow 0$ as $\vpar \rightarrow \infty$ faster than the logarithmic singularity from the $\ln f$ term. Eliminate the partial derivatives in \eqr{\ref{eq:entrEvNew}} using \eqsr{\ref{eq:gkfpo-para}}{\ref{eq:gkfpo-mu}}, resulting in
\begin{equation} \label{eq:entrD}
\pd{S}{t} = \frac{1}{\nu\vt^2}
\int_{-\infty}^{\infty} \frac{1}{\Jac}\left[ \frac{1}{\Jac f}\left(F_{\vpar}^2 + \frac{B}{2m \mu} F_{\mu}^2 \right) - \nu(\vpar-\upar)F_{\vpar} - \nu\frac{B}{m} F_\mu \right] \thinspace \dThvv.
\end{equation}
With the definitions of $F_{\vpar}$ and $F_\mu$, the second and third terms in~\ref{eq:entrD} become
\begin{equation}
\frac{1}{\nu\vt^2}\int_{-\infty}^{\infty} \nu \left[ -\left(\vpar^2 + \frac{2 B \mu}{m}\right) f + (2 \vpar \upar - \upar^2 + 3 \vt^2)f \right] \thinspace \dThvv
= 0,
\end{equation}
after integration by parts and using definitions of the moments. Therefore, since $\mu \ge 0$, \eqr{\ref{eq:entrD}} indicates that
\begin{equation}
\pfrac{S}{t} = \frac{1}{\nu \vt^2}
\int_{-\infty}^{\infty} \frac{1}{\Jac^2 f} \left(F_{\vpar}^2 + \frac{B}{2 m \mu} F_{\mu}^2 \right) \ge 0
\end{equation}
as long as $f > 0$.

The definition of entropy and the fact that it is a non-decreasing function can be used to show that the maximum entropy solution to the \mFPO~is the Maxwellian given by
\begin{equation}
    f_M(n,\upar,\vt) = \frac{n}{(2\pi \vt^2)^{3/2}}\exp\left[-\frac{(\vpar-\upar)^2+2\mu B/m}{2\vt^2}\right], \label{eq:maxwellian}
\end{equation}
where $n$ is the zeroth moment or particle number density, $n=M_0$. Such distribution arises from maximize the entropy $S$ subject to the constraint that density, momentum and energy do not change during the evolution. In other words, it is the result of finding the extrema of
\begin{equation}
    S = -\int_{-\infty}^{\infty} f \ln{f} \thinspace \dThvv
    + \lambda_0 \left(\int_{-\infty}^{\infty} \Jac f \thinspace \dThvv - M_0\right) 
    + \lambda_1 \left(\int_{-\infty}^{\infty} \vpar \Jac f \thinspace \dThvv - M_1 \right) 
    + \lambda_2 \left(\int_{-\infty}^{\infty} (\vpar^2 + 2\mu B/m) \Jac f \thinspace \dThvv - M_2\right),
\end{equation}
where $\lambda_0, \lambda_1$ and $\lambda_2$ are Lagrange multipliers. Varying this Lagrangian and applying the constraints to determine the Lagrange multipliers, leads to the Maxwellian. Because the entropy is monotonically increasing, the Maxwellian maximizes the entropy. These are textbook observations of a good collision operator, yet in section~\ref{sec:results} we will see that the meaning of a discrete maximum entropy solution must be examined carefully.

A final property of the \mFPO, which certain numerical schemes would also benefit from, is its self-adjointness. This means that for arbitrary functions $g(t,\v{R},\vpar,\mu)$, $\Jac f(t,\v{R},\vpar,\mu)$ the \mFPO~satisfies 
\begin{equation}
\inProd{g}{\Jac \gkcoll{f}} = \inProd{\Jac f}{\gkcoll{g}} \label{eq:lbo-adjoint}
\end{equation}
with the inner product defined as
\begin{equation}
\inProd{\Jac f}{g} = 
\intInf \frac{1}{f_M}\Jac f g \thinspace \dThvv
\end{equation}
where $f_M$ is the Maxwellian that satisfies $\gkcoll{f_M}=0$. Self-adjointness is demonstrated integrating \eqr{\ref{eq:lbo-adjoint}} by parts and using \ref{eq:gkfpo-para} and \ref{eq:gkfpo-mu} in order to arrive at
\begin{equation}
    \inProd{g}{\Jac \gkcoll{f}} =
    -\intInf
   \left( F_{\vpar} \pd{}{\vpar} + F_\mu \pd{}{\mu} \right) \left(\frac{g}{f_M}\right)
    \thinspace \dThvv.
\end{equation}
Then use following the identities
\begin{align}
\nu\vt^2 f_M \pd{}{\vpar}\left(\frac{\Jac f}{f_M}\right) = \nu(\vpar-\upar) \Jac f + \nu\vt^2 \pd{\Jac f}{\vpar} = F_{\vpar} \\
\nu 2\frac{m\vt^2}{B} f_M \mu \pd{}{\mu} \left(\frac{\Jac f}{f_M}\right) = \nu 2\mu \Jac f + \nu 2\frac{ m \vt^2}{B} \mu \pd{\Jac f}{\mu} = F_\mu
\end{align}
and write the ensuing equation as
\begin{equation}
      \inProd{g}{\Jac \gkcoll{f}} =
   - \nu\vt^2 \intInf  
    f_M  \left\{ \pd{}{\vpar}\left(\frac{\Jac f}{f_M}\right)\pd{}{\vpar}\left(\frac{g}{f_M}\right)
    + \frac{2m}{B} \mu \pd{}{\mu}\left(\frac{\Jac f}{f_M}\right) \pd{}{\mu}\left(\frac{g}{f_M}\right)\right\}
    \thinspace \dThvv.
    \label{eq:selfad}
\end{equation}
This is symmetric in $f$ and $g$, and the self-adjoint property follows. The self-adjoint property indicates that all eigenvalues of the operator are real and the solution is damped, a characteristic certain numerical schemes benefit from. This $1/f_M$ weighting in the definition of the inner product is standard in kinetic theory, further discussion is found in~\cite{Hakim2019}. 

%% file: scheme.tex
\section{The discrete gyroaveraged Dougherty operator} \label{sec:discrete}
This work is primarily concerned with the discontinuous Galerkin (DG) discretization of the \mFPO~in \eqr{\ref{eq:gk-boltzmann}}. The schemes are presented assuming three dimensions ($x$, $\vpar$ and $\mu$), but they can be easily extended to higher dimensions. We wish to find the numerical solution $f$ defined on a domain $\Omega \equiv \left[\xmin,\xmax\right]\times\left[\vparmin,\vparmax\right]\times\left[\mumin,\mumax\right]$ which is discretized by the structured rectangular phase-space mesh defined as $\Omega_{i,j,k} \equiv [x_{i-1/2},x_{i+1/2}] \times  [v_{\parallel,j-1/2},v_{\parallel,j+1/2}] \times [\mu_{k-1/2},\mu_{k+1/2}]$. The velocity extents of this mesh (except for $\mumin=0$) are typically far enough from zero that the distribution function $f$ has decreased by many orders of magnitude, although we will see below that for exact conservation we cannot assume it to be zero there. On each cell we select a set of orthogonal basis functions $\psi_\ell(x,\vpar,\mu)$, for $\ell=1,\ldots,\Np$, such that
\begin{equation}
    \int_{\Omega_{i,j,k}} \psi_\ell \psi_{m} \dxdvdmu = \delta_{\ell m} \frac{\Dx}{2}\frac{\Dvpar}{2}\frac{\Dmu}{2},
\end{equation}
where $\Dx$, $\Dvpar$ and $\Dmu$ are the cell lengths in each direction. In \gkeyll~we typically employ Serendipity bases constructed by choosing and orthornormalizing monomials from the polynomial space $\mathcal{V}_3^p = \{ x^l \vpar^m \mu^n \mid \mathrm{deg}_3(x^l \vpar^m \mu^n) \leq p \}$
of order $p$ and dimension $3$~\cite{Arnold:2011eu}, though the algorithm presented here is general to other orthonormal basis sets ($\mathrm{deg}_p$ refers to the sum of all monomial powers that appear superlinearly).

We build a DG scheme for the \mFPO~leveraging the concept of weak equality. This in turn yields a conservative {\mFPO} scheme that is also alias-free. This section describes weak equality, presented in~\cite{Hakim2019} and reproduced here for completeness, and its role in the recovery DG approach to second order derivatives, followed by the conservative DG discretization of the \mFPO.

\subsection{Weak equality and recovery DG} \label{sec:weakRDG}
For some (phase) space interval $I$ and some basis $\psi_k$, with $k \in\{1,\ldots,\Np\}$, spanning the function space $\mathcal{P}$, two functions $f$ and $g$ are \emph{weakly equal} if 
\begin{equation}
    \int_I (f-g) \psi_k \thinspace \dx = 0. \label{eq:weak-eq}
\end{equation}
That is, the projections of these functions on a given basis are equal. In finite-element theory and applied mathematics weak equality is referred to as weak equivalence or a weak solution to $f=g$~\cite{Brenner2002}. We denote a weak equality with $f \doteq g$, and in section~\ref{sec:results} we also describe how weak equalities lead to the proper spectral decomposition of DG signals. Although it is presented here in the context of DG, it is general to the use of finite-dimensional vector spaces.
%% There is nothing "formal" about the spectral decomposition. It is the correct decomposition. Also, there is no restriction on finite domains. AHH

%Weak equality is fundamental to our DG schemes. To begin with, 

The discrete form of a quantity expanded in the basis $\psi_k$ is given in terms of a weak definition. For example, the discrete forms of the first three moments of the distribution function are defined as
\begin{align}
    M_{0} &\doteq \int \Jac f \thinspace \dThvv \\
    M_{1} &\doteq \int \vpar\thinspace \Jac f \thinspace \dThvv \\
    M_{2} &\doteq \int (\vpar^2 + 2 B\mu / m)\thinspace \Jac f \thinspace \dThvv.
\end{align}
The primitive moments, $\upar$ and $\vt$, on the other hand must be computed using a combination of what we call weak multiplication and division in order for them to lie in $\mathcal{V}_3^p$. To illustrate these operations consider the definition of the mean velocity $\upar$ given by the relation
\begin{equation} \label{eq:weakUpar}
    \upar M_0 \doteq M_1.
\end{equation}
Using \eqr{\ref{eq:weak-eq}} and $\upar = \sum_\ell u_{\parallel\ell} \varphi_\ell(x)$ we can express this weak operation as a system of linear equations
\begin{equation}
    \sum_\ell u_{\parallel\ell} \int_I M_0 \varphi_\ell \varphi_k dx = \int_I M_1 \varphi_k dx,
\end{equation}
where $M_0$ and $M_1$ also have expansions in the configuration-space basis set, $\varphi_k(x)$. The inversion of this system to compute the $u_{\parallel\ell}$ coefficients needs to take place in each cell of the configuration-space grid and is referred to as \emph{weak-division}. Having obtained the expansion of the mean parallel velocity, one can then perform the weak multiplication $\upar M_1 \doteq K$ to obtain the kinetic energy $K$; this is needed to compute the thermal velocity via the weak analogue of \eqr{\ref{eq:vtSqEq}}:
\begin{equation} \label{eq:weakVtSq}
\upar M_1 + 3\vt^2M_0 \doteq M_2.
\end{equation}
However, weak division needs to be limited for numerical stability. As the function $M_0$ becomes too steep $\upar$ begins to diverge. In order to avoid this we limit weak division by performing cell-average division only (e.g. for $p=1$ $u_{\parallel,0}=M_{1,0}/M_{0,0}$ and $u_{\parallel,1}=0$) when $\left|M_{0,1}\right| < M_{0,0}/\sqrt{3}$. For more information see \cite{Hakim2019}.

The framework provided by weak equalities also leads to a natural formulation of recovery DG (RDG) for higher order derivatives and, more generally, recovering a continuous function from a discontinuous one. Suppose we wish to compute the second derivative $g \doteq\partial^2 f/\partial x^2 = f_{xx}$. Integration by parts in cell $I_j=[x_{j-1/2},x_{j+1/2}]$ gives
\begin{equation} \label{eq:fxxOneIBP}
g_k = \psi_k f_x\Bigg|^{x_{j+1/2}}_{x_{j-1/2}}-\int_{I_j} \pd{\psi_k}{x} f_x\thinspace\dx.
\end{equation}
Recall that $f$ has its own expansion, but since it is generally discontinuous from one cell to the next we need a way to compute its derivative at the cell boundaries. We could instead replace $f_x$ in the boundary term with $\frec_x$, where $\frec$ is the recovery polynomial constructed such that
\begin{align} \label{eq:RDGlr}
\frec &\doteq f_L\qquad x\in I_L\qquad\mathrm{on}~\mathcal{P}_L, \nonumber \\
\frec &\doteq f_R\qquad x\in I_R\qquad\mathrm{on}~\mathcal{P}_R.
\end{align}
One needs to perform two recoveries, one at $x_{j-1/2}$ and another at $x_{j+1/2}$. At $x_{j-1/2}$, $f_L$ refers to the function $f$ on the element $I_L=I_{j-1}=[x_{j-3/2},x_{j-1/2}]$, and $f_R$ is $f$ on $I_j$. Each of these is defined on the respective function spaces $\mathcal{P}_L$ and $\mathcal{P}_R$.

\begin{figure}[h]
  \centering
  \includegraphics[width=0.49\textwidth]{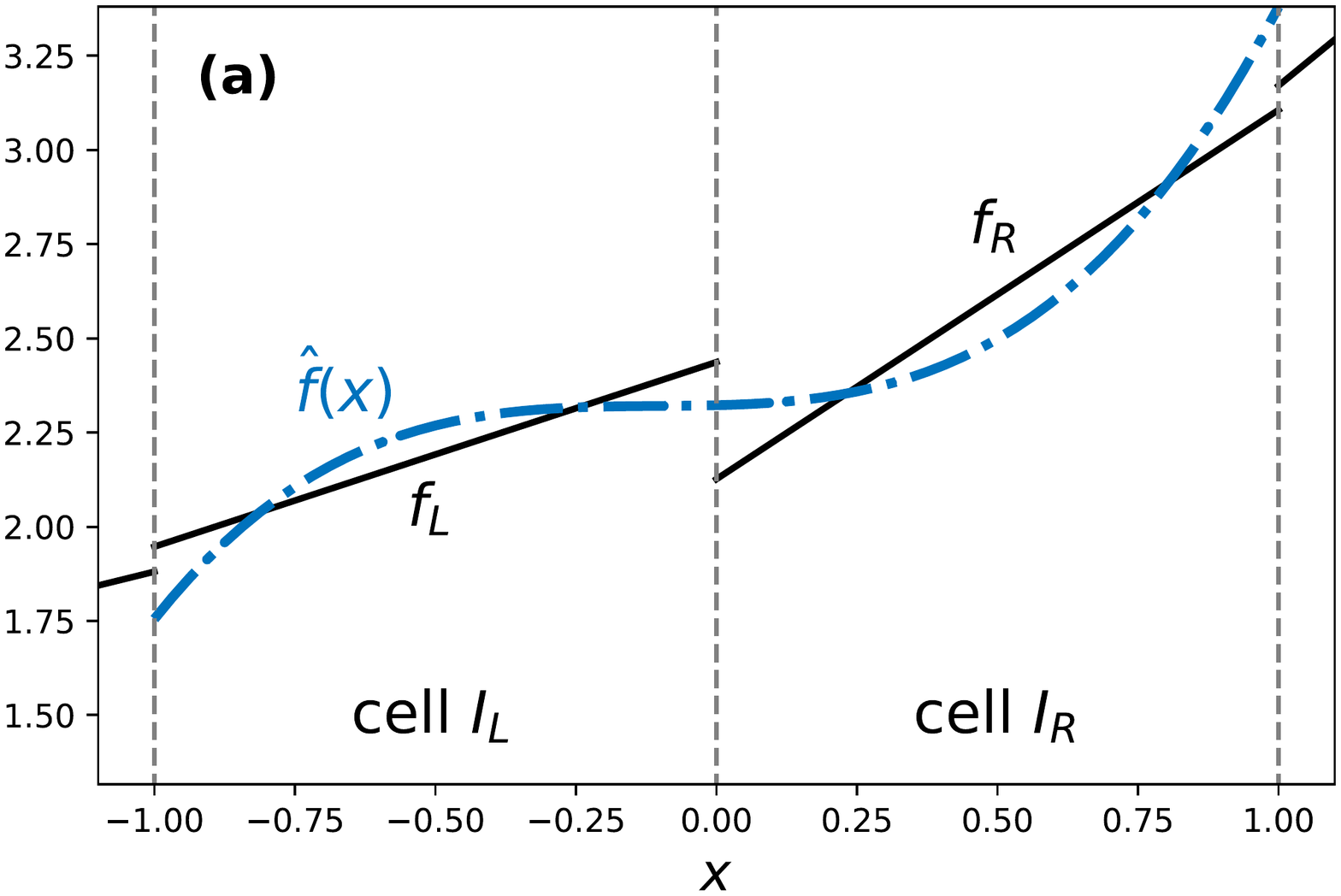}
  \includegraphics[width=0.49\textwidth]{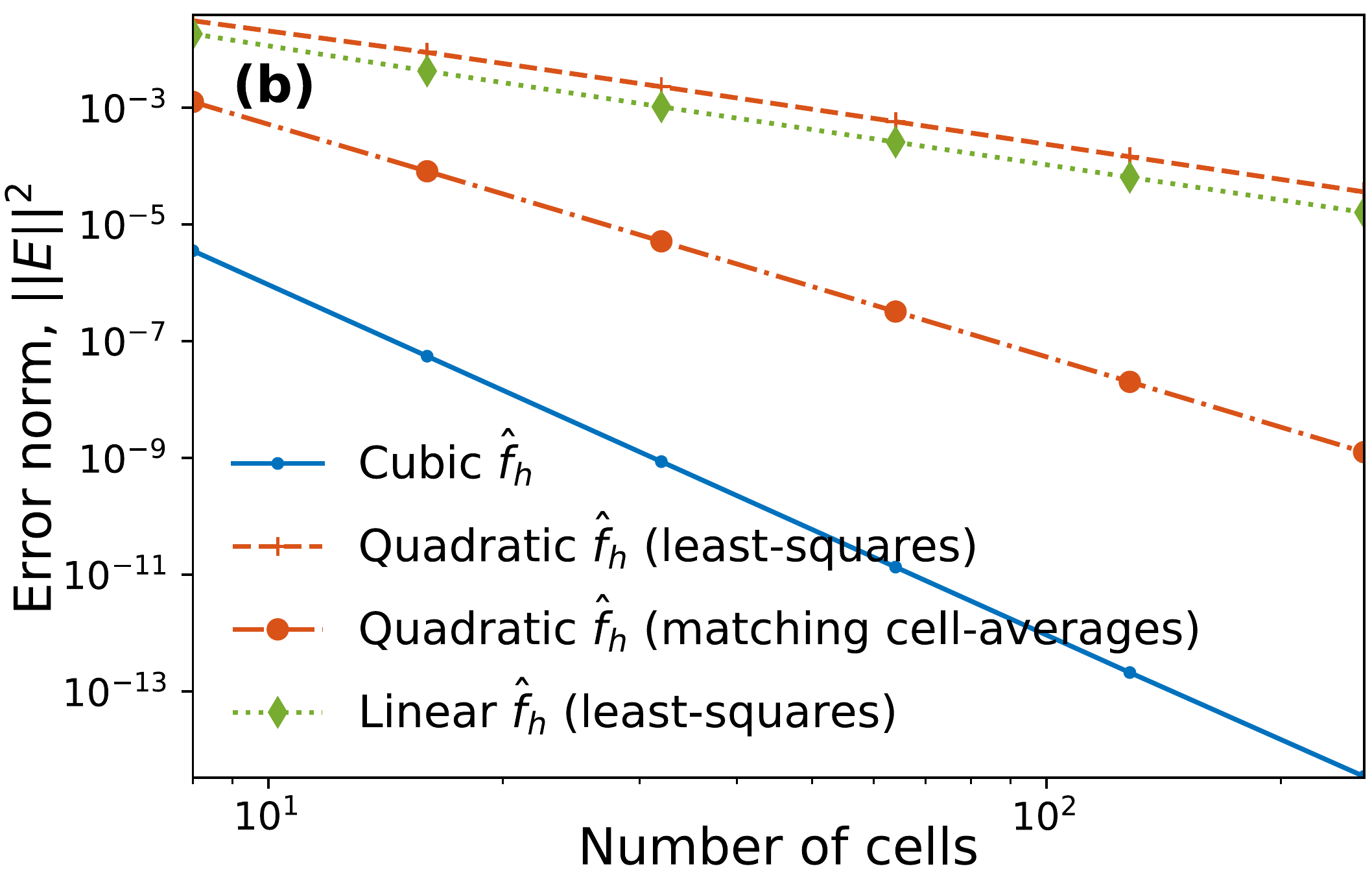}
  \caption[Recovery examples.]{(a) A DG function (solid black) and its maximal order recovery polynomial (dash-dot blue). (b) Error norm in the $d^2(\sin x)/dx^2$ with cubic, quadratic and linear recovery.}
  \label{fig:recoverySample}
\end{figure}

The equalities in \eqr{\ref{eq:RDGlr}} establish the projections of $\frec$ on $I_L$ and $I_R$, but to determine it uniquely we can use the $2\Np$ pieces of information ($\Np$ coefficients from each of $f_L$ and $f_R$) and assume $\frec$ is the maximal-order polynomial:
\begin{equation} \label{eq:recoveryMaximalOrder}
\frec(x) = \sum_{k=0}^{2\Np-1} \frec_k x^k.
\end{equation}
Replacing this definition into \eqr{\ref{eq:RDGlr}} leads to a linear system of $2\Np$ equations in the $\frec_k$ unknowns. \Figra{\ref{fig:recoverySample}}{a} illustrates an example recovery polynomial. An alternative RDG we follow here is to integrate \eqr{\ref{eq:fxxOneIBP}} a second time to arrive at
\begin{equation} \label{eq:fxxTwoIBP}
g_k = \left(\psi_k \frec_x-\pd{\psi_k}{x}\frec\right)^{x_{j+1/2}}_{x_{j-1/2}}+\int_{I_j} \pdd{\psi_k}{x} f\thinspace\dx.
\end{equation}
The system in \eqr{\ref{eq:RDGlr}} is only inverted once to obtain the ensuing stencil for $\frec$ and $\frec_x$ evaluated at the boundaries. RDG schemes of this kind were first proposed over a decade ago~\cite{VanLeer2005} as an alternative to the traditional local DG (LDG) approach to diffusion terms~\cite{Cockburn1998}. This RDG has better convergence of both cell averages and slopes upon grid refinement~\cite{Hakim2014} and leads to a conservative discrete \mFPO, which we prove in the next section.

The maximal-order recovery polynomial in \eqr{\ref{eq:recoveryMaximalOrder}} has fourth-order convergence (see \figra{\ref{fig:recoverySample}}{b}), but it does not guarantee positivity. For systems in which $f$ must remain positive, using the highest order polynomial possible may lead to incursions below zero when $f$ is small. An example of this, due to both diffusion and advection, is illustrated in section~\ref{sec:results}. We explored computing second derivatives with lower order polynomials in order to see if positivity problems are minimized. The test function $f{=}\sin x$ in $x\in[-1,1]$ was discretized with piecewise polynomial bases ($p=1,~\Np=2$) which allows for a cubic maximal-order polynomial. We could also, instead of using \eqr{\ref{eq:recoveryMaximalOrder}}, request a quadratic or a linear $\frec$. In these two cases \eqr{\ref{eq:RDGlr}} leads to an over determined system that we solved by least squares. The error norms in the computation of $d^2(\sin x)/dx^2$ for all three recoveries are given in \figra{\ref{fig:recoverySample}}{b} as a function of resolution. Unfortunately the convergence of these lower-order, least squares methods is inferior (closer to second order) and, in advection-diffusion problems, they were less stable. Also note that the quadratic least-squares $\frec$ (dashed orange in \figra{\ref{fig:recoverySample}}{b}) did not do any better than the linear $\frec$ (dotted green in \figra{\ref{fig:recoverySample}}{b}); a better procedure is to construct $\frec$ by matching the cell-averages in each cell, and seeking the least-squares solution that tries to match the slopes in both cells (orange dash-dot in \figra{\ref{fig:recoverySample}}{b}). In what follows, we use the maximal-order recovery polynomial and take other measures to decrease the likelihood of $f<0$.

\subsection{Discrete \mFPO~scheme} \label{sec:discScheme}
In order to discretize the \mFPO~we project \eqr{\ref{eq:GkFP}} onto the phase-space basis by multiplying by a test function $w\in\mathcal{V}_3^p$ and integrating over all phase-space:
\begin{equation}
\int_{\Omega_{i,j,k}} w \pd{\Jac f}{t} \dxdvdmu = \nu \int_{\Omega_{i,j,k}} w \left\lbrace \pd{}{\vpar}\left[(\vpar - \upar) \Jac f + \vt^2 \pd{\Jac f}{\vpar}\right] 
+ \pd{}{\mu}\left[2 \mu \Jac f +
2\frac{m \vt^2}{B}\mu \pd{\Jac f}{\mu} \right] \right\rbrace  \dxdvdmu,
\end{equation}
where we neglect the $2 \pi/m$ integration factor to simplify notation. Integrate by parts once to give
\begin{equation}
\begin{aligned}
\int_{\Omega_{i,j,k}} w \pd{\Jac f}{t} \dxdvdmu
&= \nu\int_{x_{i-1/2}}^{x_{i+1/2}} \int_{\mu_{k-1/2}}^{\mu_{k+1/2}} w \,
G_{\vpar}(f_L,f_R)\Bigg|_{v_{\parallel,j-1/2}}^{v_{\parallel,j+1/2}}\thinspace\dxdmu 
+ \nu\int_{x_{i-1/2}}^{x_{i+1/2}} \int_{v_{\parallel,j-1/2}}^{v_{\parallel,j+1/2}} w \,
G_{\mu}(f_L,f_R) \Bigg|_{\mu_{k-1/2}}^{\mu_{k+1/2}}\thinspace\dxdv  \\
&\quad-\nu \int_{\Omega_{i,j,k}} \left\{ \pd{w}{\vpar} 
\left[ (\vpar-\upar)\Jac f + \vt^2 \pd{\Jac f}{\vpar} \right] + \pd{w}{\mu}\left[2 \mu \Jac f + \frac{2 m \vt^2}{B} \mu \pd{\Jac f}{\mu}\right] \right\} \dxdvdmu.    \label{eq:schemenc}
\end{aligned}
\end{equation}
The quantities $G_{\vpar}(f_L,f_R)$ and $G_{\mu}(f_L,f_R)$ are numerical fluxes chosen to preserve, in addition to conservation, other properties like stability and positivity. For example, the Lax-Friedrichs (LF) penalty fluxes are
\begin{align} \label{eq:flux}
G_{\vpar}(f_L,f_R) &= \frac{1}{2}(\vpar-\upar)(\Jac f_R+\Jac f_L) - \frac{\tau_{\vpar}}{2}(\Jac f_L-\Jac f_R)+ \vt^2 \pd{\Jac \frec}{\vpar} \\
G_\mu(f_L,f_R) &= \mu(\Jac f_R+\Jac f_L) - \frac{\tau_\mu}{2}(\Jac f_L-\Jac f_R) + \frac{2 m \vt^2}{B} \mu \pd{\Jac \frec}{\mu} \nonumber, 
\end{align}
where $\tau_{\vpar}= \max(|\vpar-\upar|)$ and $\tau_{\mu} = \max(2\mu)$. The notation $f_L$ and $f_R$ refer to the distribution function in the left and right cells of a boundary, respectively. In section~\ref{sec:stability} we also discuss pure upwind fluxes which are more beneficial for positivity. Notice that the contribution to the numerical flux arising from the diffusion term is computed using the recovery polynomial described in section~\ref{sec:weakRDG}.

Additionally we impose the following boundary conditions on the numerical fluxes:
\begin{equation} \label{eq:fluxbc}
\begin{aligned}
    G_{\vpar}\left(f_L(\vparmin),f_R(\vparmin)\right) &= G_{\vpar}\left(f_L(\vparmax),f_R(\vparmax)\right) = 0, \\
    G_{\mu}\left(f_L(0),f_R(0)\right) &= G_\mu \left(f_L(\mu_{\max}),f_R(\mu_{\max})\right) = 0.
\end{aligned}
\end{equation}
The discrete form in \eqsr{\ref{eq:schemenc}}{\ref{eq:fluxbc}} does not conserve momentum, shown by substituting $w=\vpar$ into \eqr{\ref{eq:schemenc}}~and summing over $\vpar$ space ($j$ index). The diffusion term in the volume integral can be integrated by parts again, yielding a surface term with non-vanishing jumps that breaks momentum conservation. A similar issue arises with energy conservation.

The conservation of $M_0$, $M_1$ and $M_2$ can be guaranteed if one integrates by parts twice and evaluates the additional surface terms using the recovered distribution function $\frec$. Therefore the conservative \mFPO~DG scheme follows
\begin{equation}
\begin{aligned}
\int_{\Omega_{i,j,k}} w \pd{\Jac f}{t} \dxdvdmu
&=
\nu\int_{x_{i-1/2}}^{x_{i+1/2}} \int_{\mu_{k-1/2}}^{\mu_{k+1/2}} \left(w \,G_{\vpar}(f_L,f_R)- \pd{w}{\vpar} \vt^2 \Jac \frec\right)_{v_{\parallel,j-1/2}}^{v_{\parallel,j+1/2}} \thinspace\dxdmu \\
&\quad + \nu\int_{x_{i-1/2}}^{x_{i+1/2}} \int_{v_{\parallel,j-1/2}}^{v_{\parallel,j+1/2}} \left(w \,
G_{\mu}(f_L,f_R)- \pd{w}{\mu} 2\frac{m \vt^2}{B} \mu\Jac\frec\right)_{\mu_{k-1/2}}^{\mu_{k+1/2}}\thinspace\dxdv \\
&\quad - \nu\int_{\Omega_{i,j,k}} \left\{ \pd{w}{\vpar}
 (\vpar-\upar) - \pdd{w}{\vpar}\vt^2 + \pd{w}{\mu}2 \mu - 2\frac{ m \vt^2}{B} \left(\mu \pdd{w}{\mu} + \pd{w}{\mu}\right) \right\}\Jac f \, \dxdvdmu.    \label{eq:schemec}
\end{aligned}
\end{equation}
We continue to assume a $2 \pi/m$ factor in front of the $\mu$ integral. The conservative properties of the discrete operator are proven below.

\subsubsection{Number Density Conservation:}

Scheme~\ref{eq:schemec} conserves number density:
\begin{equation}
    \frac{d}{d t} \sum_{j,k}  \int_{\Omega_{i,j,k}} \Jac f \dxdvdmu = 0.
\end{equation}
In order to show this use $w=1$ in \eqr{\ref{eq:schemec}} and sum over all velocity space cells. The sum need not be over configuration space as the configuration space gradients only occur in the collisionless terms of the gyrokinetic equation. The numerical flux is continuous across interior cell surfaces, so those contributions cancel. Only global the fluxes at the boundaries of velocity space remain, but those are zero given the boundary conditions in \eqr{\ref{eq:fluxbc}}.

\subsubsection{Discrete Momentum Conservation:} \label{sec:discM1con}

Scheme~\ref{eq:schemec} conserves momentum:
\begin{equation}
    \frac{d}{d t} \sum_{j,k}  \int_{\Omega_{i,j,k}} \vpar \Jac f \dxdvdmu = 0,
\end{equation}
if the following \textit{weak-equality relation} is satisfied:
\begin{equation}
    \upar M_0- \vt^2 \sum_k \int_{\mu_{k-1/2}}^{\mu_{k+1/2}} \left(\Jac f(\vparmax)-\Jac f(\vparmin)\right) \thinspace \dmu \doteq M_{1}. \label{eq:momWeakConst}
\end{equation}
One can show momentum conservation and arrive at this constraint using $w=\vpar$ in \eqr{\ref{eq:schemec}} and summing over all velocity space cells to get
\begin{equation}
\begin{aligned}
\frac{d}{d t} \sum_{j,k}  \int_{\Omega_{i,j,k}} \vpar \Jac f \dxdvdmu
    = -\nu\sum_{j,k} \int_{x_{i-1/2}}^{x_{i+1/2}} \int_{\mu_{k-1/2}}^{\mu_{k+1/2}} \vt^2 \Jac \frec \dxdmu \Bigg|_{v_{\parallel,j-1/2}}^{v_{\parallel,j+1/2}}
    -\nu\sum_{j,k}
\int_{\Omega_{i,j,k}} (\vpar-\upar)\Jac f \, \dxdvdmu
\end{aligned}
\end{equation}
The contributions from the numerical fluxes $G_{\vpar}$  and $G_\mu$ drop out due to continuity and boundary conditions. In the first term all interface contributions from $\vpar$ will cancel except the first and last. Combined with the definition of the discrete moments in the second term leads to the constraint
\begin{equation}
    \int_{x_{i-1/2}}^{x_{i+1/2}}
    \left[ \sum_k \int_{\mu_{k-1/2}}^{\mu_{k+1/2}}
    \vt^2\left(\Jac\frec(\vparmax) - \Jac\frec(\vparmin) \right)\thinspace\dmu
    + M_1 - \upar M_0 \right] \thinspace \dx
    = 0. \label{eq:momConst}
\end{equation}
Using the definition of weak equality, this implies that the momentum will be conserved if \eqr{\ref{eq:momWeakConst}} is satisfied. Notice that $\frec$ was replaced by $f$ because at the outer velocity boundaries there is no ``outside'' cell to allow for recovery of a continuous distribution function.

The weak-equality constraint \eqr{\ref{eq:momWeakConst}} is stronger than what is required by \eqr{\ref{eq:momConst}}. However, ensuring that the weak-equality constraint is satisfied automatically ensures that momentum conservation is preserved. For simplicity we assume a unit mass $m=1$ in what follows, with no loss of generality.

\subsubsection{Discrete Energy Conservation:} \label{sec:discM2con}

Scheme~\ref{eq:schemec} conserves energy
\begin{equation}
    \frac{d}{d t} \sum_{j,k}  \int_{\Omega_{i,j,k}} \left(\frac{1}{2} \vpar^2 + \mu B \right)\Jac f \dxdvdmu = 0,
\end{equation}
if the following weak-equality relation is satisfied:
\begin{equation}
\begin{aligned}
    \upar M_1 + \vt^2 &\left[ 3 M_0 - \sum_k \int_{\mu_{k-1/2}}^{\mu_{k+1/2}} \left(\vparmax \Jac f(\vparmax)  - \vparmin \Jac f(\vparmin) \right)\thinspace\dmu \right. \\
    &\left.\quad\quad\hspace{3pt}- \sum_j \int_{v_{\parallel,j-1/2}}^{v_{\parallel,j+1/2}} 2 \left(\mumax \Jac f(\mumax) - \mumin \Jac f(\mumin) \right)\thinspace\dvpar \right] \doteq M_2. \label{eq:erWeakConst}
\end{aligned}
\end{equation}
Assuming that $p\ge 2$ such that $v^2 \in \mathcal{V}_2^p$, the above two equations follow from replacing $w=\vpar^2/2 + \mu B$ in \eqr{\ref{eq:schemec}} and summing over all velocity space cells, which yields
\begin{equation}
\begin{aligned}
    \frac{d}{d t} \sum_{j,k}  \int_{\Omega_{i,j,k}} \frac{1}{2} v^2 \Jac f \dxdvdmu
    &= -\nu\sum_{j,k} \int_{x_{i-1/2}}^{x_{i+1/2}} \left(\int_{\mu_{k-1/2}}^{\mu_{k+1/2}} \vt^2 \vpar \Jac\frec \thinspace\Bigg|_{v_{\parallel,j-1/2}}^{v_{\parallel,j+1/2}}\thinspace\dmu 
    + \int_{v_{\parallel,j-1/2}}^{v_{\parallel,j+1/2}}  2 \vt^2 \, \mu\Jac\frec \thinspace\Bigg|_{\mu_{k-1/2}}^{\mu_{k+1/2}}\thinspace\dvpar\right)\thinspace\dx& \nonumber \\
    &\quad- \nu\sum_{j,k} \int_{\Omega_{i,j,k}} \left[ \vpar (\vpar-\upar) + 2 \mu B - 3\vt^2  \right]\Jac f \dxdvdmu.
\end{aligned}
\end{equation}
All contributions from interior cell interfaces cancel in the first term. Velocity-space integrals in the second term can be written in terms of discrete moments, leading to the following constraint in order to have energy conservation:
\begin{equation} \label{eq:erConst}
\begin{aligned}
    &\int_{x_{i-1/2}}^{x_{i+1/2}}
    \left[
    \sum_k \int_{\mu_{k-1/2}}^{\mu_{k+1/2}} \vt^2\left(\vparmax\Jac\frec(\vparmax) - \vparmin\Jac\frec(\vparmin) \right) \thinspace\dmu  \right. \\
    &\left.\quad\qquad+ \sum_j \int_{v_{\parallel,j-1/2}}^{v_{\parallel,j+1/2}} 2 \vt^2 \left(\mumax\Jac\frec(\mumax)-\mumin\Jac\frec(\mumin)\right)\thinspace\dvpar+ M_2 - \upar M_1 - 3\vt^2 M_0
    \right] \thinspace\dx
    = 0.
\end{aligned}
\end{equation}
Using the definition of weak equality this implies that the energy will be conserved if \eqr{\ref{eq:erWeakConst}} is satisfied.

Thus, exact $p\geq2$ conservation of momentum and energy leads to the following set of weak-equality relations:
\begin{equation}
\begin{aligned}
\upar M_0 - \vt^2 \frac{2\pi}{m}\sum_k \int_{\mu_{k-1/2}}^{\mu_{k+1/2}} \Jac f\thinspace\Big|^{\vparmax}_{\vparmin} \thinspace \dmu \doteq M_1, \\
\upar M_1 + \vt^2 \left[ 3 M_0 - \frac{2\pi}{m}\sum_k \int_{\mu_{k-1/2}}^{\mu_{k+1/2}} \vpar \Jac f\thinspace\Big|^{\vparmax}_{\vparmin}\thinspace\dmu - \frac{2\pi}{m}\sum_j \int_{v_{\parallel,j-1/2}}^{v_{\parallel,j+1/2}} 2 \mu \Jac f\thinspace\Big|^{\mumax}_{\mumin} \thinspace\dvpar \right] \doteq M_2,
\end{aligned}
\end{equation}
where we restored the $2\pi/m$ factors for clarity.
This is a weak linear system of equations that needs to be inverted in every cell of the configuration-space grid to compute the parallel drift velocity $\upar$ and the thermal speed $\vt$. Without the boundary corrections presented above, the errors in the conserved quantities are several orders of magnitude higher. Also, instabilities can be observed when velocity-grid extents are too low and $f$ is appreciable at the boundary.

\subsubsection{Discrete p=1 Energy Conservation:} \label{sec:discM2conP1}

The above energy conservation theorem applied to $p \ge 2$ basis functions which span the quadratic term in the test function $\vpar^2/2 + \mu B$. For a piecewise linear basis ($p=1$), conservation can be maintained in the sense that we can conserve the projection of the second moment $M_2$ onto the piecewise linear basis. This property can be ensured if the quadratic term in the test function is replaced by its projection onto the basis functions, $\overline{\vpar^2} \in \mathcal{V}_3^1$, which is weakly equivalent to $\vpar^2$ in this basis: 
\begin{equation}
    \overline{\vpar^2} \doteq \vpar^2 \quad \mathrm{on} \; \mathcal{V}_3^1. \label{eq:v2bar}
\end{equation}
It is straightforward to show that $\overline{\vpar^2}$ is continuous. Also, in a weak-equality sense, the definition of particle energy is the same, whether we use the original quadratic expression or its projection. 
%\begin{equation}
%    E = \int_{-\vparmax}^{\vparmax} \int_{\mumin}^{\mumax} \, (\vpar^2/2 + \mu B) \thinspace\Jac f \dvdmu
%\end{equation}
Therefore we can show that scheme~\ref{eq:schemec} satisfies
\begin{equation} \label{eq:enerCon}
    \d{}{t} \sum_{j,k}  \int_{\Omega_{i,j,}} (\vpar^2/2 + \mu B)\thinspace\Jac f \dxdvdmu = 0,
\end{equation}
with a piecewise linear basis as long as the following weak equality is satisfied:
\begin{equation} \label{eq:erWeakConstP1}
\begin{aligned}
    \upar M^*_1 +\vt^2&\left[M^*_0 + 2 M_0 -\sum_k \int_{\mu_{k-1/2}}^{\mu_{k+1/2}} \left( \vcen_{\parallel,\max}\Jac f(\vparmax)-\vcen_{\parallel,\min} \Jac f(\vparmin)\right) \dmu \right. \\ 
    &\left.\qquad\qquad- \sum_j \int_{v_{\parallel,j-1/2}}^{v_{\parallel,j+1/2}} 2 \left(\mumax\Jac f(\mumax)-\mumin\Jac f(\mumin)\right)\dvpar\right]  \doteq M^*_2.
\end{aligned}
\end{equation}
where $\vcen_{\parallel,j} = (v_{\parallel,j+1/2}+v_{\parallel,j-1/2})/2$, $\Delta v_{\parallel,j} = \vcen_{\parallel,j+1}-\vcen_{\parallel,j}$ and the ``star moments'' are defined as
\begin{equation} \label{eq:starMoms}
\begin{aligned}
    M^*_0 &\doteq \sum_{j \neq j_{\max}} \sum_k \int_{\mu_{k-1/2}}^{\mu_{k+1/2}} \Delta v_{\parallel,j} \thinspace \Jac\frec_{j+1/2} \, \dmu \\
    M^*_1 &\doteq \sum_{j,k} \int_{v_{\parallel,j-1/2}}^{v_{\parallel,j+1/2}} \int_{\mu_{k-1/2}}^{\mu_{k+1/2}} \vcen_{\parallel,j} \thinspace \Jac f \, \dvdmu\\
    M^*_2 &\doteq \sum_{j,k} \int_{v_{\parallel,j-1/2}}^{v_{\parallel,j+1/2}} \int_{\mu_{k-1/2}}^{\mu_{k+1/2}} (\vcen_{\parallel,j} \vpar + 2\mu B) \thinspace \Jac f \thinspace \dvdmu.
\end{aligned}
\end{equation}

The argument leading to~\ref{eq:enerCon}-\ref{eq:starMoms} begins with using the projection $\overline{\vpar^2}/2$ (\ref{eq:v2bar}) in showing conservation of the energy, since $\vpar^2/2\notin\mathcal{V}_3^1$. This means setting $w = \overline{\vpar^2}/2 + \mu B$ in \eqr{\ref{eq:schemec}} and summing over all velocity space cells. Aided by the fact that
\begin{equation}
    \pd{}{\vpar} \left(\frac{1}{2} {\overline{\vpar^2}} \right) = \frac{1}{2}(v_{\parallel,j+1/2} + v_{\parallel,j-1/2}) \equiv \vcen_{\parallel,j}.
\end{equation}
one then arrives at
\begin{equation} \label{eq:energyp1}
\begin{aligned}
    \frac{d}{d t} \sum_{j,k} \int_{\Omega_{i,j,k}} (\vpar^2/2 + \mu B) \thinspace \Jac f \dxdvdmu
    &= -\nu\sum_{j,k} \int_{x_{i-1/2}}^{x_{i+1/2}}\left( \int_{\mu_{k-1/2}}^{\mu_{k+1/2}} \vcen_{\parallel,j} \vt^2 \Jac\frec \thinspace \Bigg|_{v_{\parallel,j-1/2}}^{v_{\parallel,j+1/2}} \dmu 
    + \int_{v_{\parallel,j-1/2}}^{v_{\parallel,j+1/2}}  2 \vt^2 \, \mu \Jac\frec \,
    \Bigg|_{\mu_{k-1/2}}^{\mu_{k+1/2}}\dvpar\right)\dx \\
    &\quad-\nu\sum_{i,j} \int_{\Omega_{i,j,k}} \left[ \vcen_{\parallel,j} (\vpar-\upar) + 2\mu B  - 2\vt^2 \right] \Jac f\dxdvdmu.
\end{aligned}
\end{equation}
The terms containing the numerical flux $G_{\vpar}$ drop out since $\overline{\vpar^2}/2$ is continuous and we are enforcing zero-flux boundary conditions in velocity-space (and so does the $G_{\mu}$ term). However, as $\vcen_{\parallel,j}$ is not continuous the contribution from the first term in \eqr{\ref{eq:energyp1}} does not drop out. This term can be written as
\begin{equation}
\begin{aligned}
    \int_{x_{i-1/2}}^{x_{i+1/2}} \vt^2 \sum_{j,k} \int_{\mu_{k-1/2}}^{\mu_{k+1/2}} \vcen_{\parallel,j} \Jac \frec \Bigg|_{v_{\parallel,j-1/2}}^{v_{\parallel,j+1/2}} \thinspace \dxdmu 
    = \int_{x_{i-1/2}}^{x_{i+1/2}} \sum_k \int_{\mu_{k-1/2}}^{\mu_{k+1/2}}
    &\left[
    \vt^2 \left( \vcen_{\parallel,\max} \Jac f(\vparmax)-\vcen_{\parallel,\min} \Jac f(\vparmin)\right) \right. \\
    &\left.\quad- \vt^2 \sum_{j \neq j_{\max}} \Delta v_{\parallel,j} \Jac \frec_{j+1/2}
    \right] \thinspace \dxdmu.
\end{aligned}
\end{equation}
Utilizing the star moments in \eqr{\ref{eq:starMoms}} and the definition of weak equality, this last relation implies that the energy will be conserved in the $p=1$ case if \eqr{\ref{eq:erWeakConstP1}} is satisfied.

In summary, for piecewise linear bases the drift velocity and thermal speed must be determined using the following set of linear weak-equality relations
\begin{equation}
\begin{aligned}
\upar M_0 &- \vt^2 \frac{2\pi}{m}\sum_k \int_{\mu_{k-1/2}}^{\mu_{k+1/2}} \Jac f\thinspace\Big|^{\vparmax}_{\vparmin} \thinspace \dmu \doteq M_1, \\
\upar M^*_1 &+\vt^2\left\lbrace M^*_0 + 2 M_0 -\frac{2\pi}{m}\left[\sum_k \int_{\mu_{k-1/2}}^{\mu_{k+1/2}} \left( \vcen_{\parallel,\max}\Jac f(\vparmax)-\vcen_{\parallel,\min} \Jac f(\vparmin)\right) \dmu  \right.\right. \\
&\left.\left.\qquad\qquad\qquad\qquad\quad+ 2\sum_j \int_{v_{\parallel,j-1/2}}^{v_{\parallel,j+1/2}}  \mu \Jac f\thinspace\Big|^{\mumax}_{\mumin}\dvpar\right]\right\rbrace  \doteq M^*_2.
\end{aligned}
\end{equation}
We have again reinstated the $2\pi/m$ factors for completeness. Notice that this weak system requires computing the first two regular moments ($M_0$ and $M_1$) and the three star moments.

%% file: stability.tex
\section{Time-stepping and stability} \label{sec:stability}
A high-order, conservative DG scheme for the \mFPO~must be accompanied by a suitable time-stepping scheme. In this section we complement the spatial discretization of the gyrokinetic-\mFPO~equation presented in sections~\ref{sec:theory}-\ref{sec:discrete}, and that of the Vlasov-Maxwell-Dougherty~system presented in~\cite{Hakim2019}, with a description of the time-stepping algorithm and its stability. As a preliminary, recall that one can determine the appropriate time-step ($\Dt$) for a linear problem $df/dt = \mathrm{L}[f]$ by estimating the eigenvalues $\eig$ of the operator $\mathrm{L}$. Then the time-step is chosen such that $\eig \Dt$ is within the region of numerical stability for a particular time-stepping algorithm for all eigenvalues of the operator~\cite{Durran2010}.

Purely damped modes, those for which $df/dt =  \eig f$ and $\eig < 0$, 
% gwh: if they are purely damped, there is
% no imaginary part.
will be stable when using an individual Euler step if $\left|\eig\right|\Dt<2$ because $f^{n+1}=(1+\eig\Dt)f^n$ (where the $n$ exponent labels the $n$-th time step). In \gkeyll~we instead use an explicit third-order Strong Stability Preserving (SSP) Runge-Kutta (SSP-RK3) with convex combinations of individual Euler steps that has a combined stability limit of $\left|\eigmax\right|\Dt \lesssim 2.512$ for purely damped modes,
as can also be seen in figure 2.4 in \cite{Durran2010}.
% gwh: the previously cited limit of
% \eigmax \Dt < 1.73 is incorrect, that is the
% limit for purely oscillatory modes.
% I cite this number of 2.512 to such high
% precision for later comparison with a paper
% by Cockburn and Shu.
% 
% It is not hard to find this 2.512 yourself. The amplification factor for RK3 is
% A = 1 + s + s^2/2 + s^3/6
% where s = lambda*dt.  
% (For RK methods where the number of stages N is identical to the order of the method, the amplification factor is identical to an Nth order Taylor-Series expansion of the exact solution exp(s).
%
% Just plot this vs. s
% and find where it crosses A = -1.
%
One may instead wish to ensure the stability of each Euler stage and avoid ``overdamped" solutions that oscillate around zero (instead of just damping with the same sign) by using the more conservative limit $\left|\eig\right|\Dt<1$.

Although the \mFPO~is nonlinear we can use these ideas to estimate stability limits of the drag and diffusion terms separately. We will then combine these into a single rule for choosing the time step for the whole \mFPO, and discuss the additional considerations brought about by the collisionless terms. We add that for collision-dominated regimes explicit time-stepping will not be sufficiently efficient, and one may need to use an implicit scheme instead. One option we intend to explore in the future is to use super-time-steppers~\cite{Meyer2014} to overcome this obstacle.

\subsection{Stability of a DG Advection Operator}

The stability limits of the \mFPO~drag term can be probed by examining the advection equation $\partial f/\partial t = -v\partial f/\partial x$. The lowest order ($p=0$) DG is equivalent to a first-order upwind finite volume algorithm, which for $v>0$ gives the solution in the $j$-th cell as
\begin{equation}
\frac{\partial f_j}{\partial t} = - v \frac{f_j - f_{j-1}}{\Dx}.
\end{equation}
One can substitute the eigenfunction $f_j(t) = A(t) {\rm e}^{i k x_j}$ and find that the maximum eigenvalue occurs for $k=\pi/\Delta x$, the Nyquist mode. This simplifies the above equation to
$\partial A/\partial t = -2vA/\Dx$ and hence the largest absolute magnitude of the eigenvalue is
\begin{equation}
\eigmax = \frac{2 v}{\Delta x} \quad\mathrm{for}~p=0.
\label{eq:lambda_DG1}
\end{equation}
As mentioned earlier the stability limit of a first-order Euler step is $|\eig|\Dt<2$ (see figure 2.4 in~\cite{Durran2010}). Using \eqr{\ref{eq:lambda_DG1}}, this condition can be expressed in terms of a Courant-Lewy-Friedrichs ($\cfl$) number as $v\Dt/\Dx=\cfl<1$. Requiring every Euler stage of a SSP-RK3 method to satisfy this constraint can be overly conservative, and one could, in principle, use a slightly larger time-step, as described above, $\eigmax \Delta t < 2.512$, which corresponds to $\cfl < 1.256$.

For piecewise linear DG ($p=1$), the maximum eigenvalue is somewhat larger. In general one can do a von Neuman-type analysis assuming the solutions between cells varies like $e^{i k x}$, and calculate the spectrum of eigenvalues of the DG discretization of the $d/dx$ operator as a function of wavenumber $k$, from $k=0$ up to $k=k_{\rm max} = (p+1)\pi/(\Delta x)_{\rm cell}$, where the extended $k$ domain accounts for the $(p+1)$ degrees of freedom within each DG cell that effectively provide a finer mesh than the cell width.
We will assume, as is true for $p=2$, that the  stability limit is set by the Nyquist mode, $k=k_{\rm max}$. 
% Interpreting the $p+1$ degrees-of-freedom within a cell in high-order DG to correspond to the smaller effective cell length $\Dx=\Dxcell/(p+1)$, the Nyquist wavenumber would be $k=\pi(p+1)/\Dxcell$. Actually, as we will see shortly, hyperbolic terms behave as if $\Dx=\Dxcell/(2p+1)$. 
In the $j$-th cell the Nyquist mode for $p=1$ has a mean value of 0 (so $f_{j0}=0$) and a linear slope $f_j(x,t)=f_{j1}(t)\psi_1(x-x_j)$, where $\psi_1(x)=\sqrt{3}x/(\Dx/2)$ is an orthonormal basis function. Then the evolution of the DG representation of this mode is
\begin{equation}
\begin{aligned}
\frac{\partial f_1}{\partial t} & = - v \, \inProd{\psi_1}{\frac{\partial f}{\partial x}} = - \frac{v \sqrt{3}}{\Dx} \big(\hat{F}(\Dx/2) + \hat{F}(-\Dx / 2) \big) \\
& = - 6 \frac{v}{\Dx}f_1,
\end{aligned}
\end{equation}
where $v \hat{F}$ is the upwind numerical flux at the cell boundaries. In this case, the eigenvalue is
\begin{equation}
\lambda_{\rm max} = 6 \frac{v}{\Delta x}_{\rm cell}.
\label{eq:lambda_DG2}
\end{equation}
Equations \ref{eq:lambda_DG1} and \ref{eq:lambda_DG2} are fit perfectly by the formula $\eigmax = 2(2 p +1)v / \Dxcell$. However, for piecewise quadratic basis functions ($p=2$), one finds that $\eigmax \approx 11.9 \, v / \Dxcell$. (This is done by solving for the $\eig$ spectrum of the DG discretization of $\partial/\partial x$ which, for $p=2$, requires solving a $3 \times 3$ linear system.) The resulting general formula for the maximum eigenvalue for advection is
\begin{equation}
    \lambda_{\adv} = 2 \, C_{\adv,p} \, (2 p + 1) \max(v/\Dxcell), \label{eq:advec-cell}
\end{equation}
where the advection coefficient is $C_{\adv,p}=\{1,1,1.2\}$ for $p=\{0,1,2\}$, respectively.
For RK3 $\lambda_{\rm max} \Delta t < 2.512$ gives a time step limit of $v \Delta t / \Delta x_{\rm cell} < \{1.256, \, 0.418, \, 0.209\}$ for $p=\{0,1,2\}$, which is within 3\% of the empirically determined numbers in Table 2.2 of \cite{Cockburn2001flux}.  Note that an equivalent finite-difference/finite-volume mesh with the same number of degrees of freedom would have an effective grid spacing $\Delta x_{\rm eff} = \Delta x_{\rm cell}/(p+1)$, so the stability limit on the time step in terms of an effective Courant number is $v \Delta t / \Delta x_{\rm eff} < 1.256 \, (p+1) / ( C_{adv,p} (2 p + 1)) = \{1.256, \, 0.836, \, 0.627\}$ for $p=\{0,1,2\}$, which does not drop as quickly at higher $p$ as a Courant number $v \Delta t / \Delta x_{\rm cell}$ expressed in terms of cell width.

\subsection{Stability of a DG diffusion operator}
Consider the diffusion equation $\partial f / \partial t = D \partial^2 f/ \partial x^2$. In second-order centered finite-difference (equivalent to finite-volume) discretizations of this operator the largest magnitude eigenvalue is $\eigmax= -4D/ (\Dx)^2$ when using a forward Euler step. This is the $p=0$ limit of DG, and one might attempt to employ the same formula but with the effective DG cell length $\Dx=\Dxcell/(2p+1)$ that was used for advective terms. For a parabolic term (diffusion) it turns out that $\Dx = \Dxcell/(p +1)$ is more accurate. van Leer and Nomura calculate the eigenmodes and eigenvalues of a diffusion operator using RDG with $p=1$ and $p=2$~\cite{VanLeer2005}.
Their equation 79 and figure 1 give $\eigmax = 15 / (\Dxcell)^2$ for $p=1$, while their figure 3 gives $\eigmax \approx 33 / (\Dxcell)^2$ for $p=2$. 
These results can be fit with the expression
\begin{equation}
\eig_d = - 4C_{\dif,p} D \left(\frac{p+1}{\Dxcell}\right)^2.
\label{eq:diff-cell}
\end{equation}
where $C_{\dif,p}=\{1,0.94,0.92\}$ for $p=\{0,1,2\}$, respectively.

%There are some subtleties regarding interpretation of the results. While we agree with their calculation of the spectrum of eigenvalues, we have a different interpretation of the multiple eigenvalue roots, as we interpret them as all being physical. For example, their Fig.\ 3 shows that the second branch is properly interpreted as giving the physical eigenvalues for the range $\beta = k \Delta x = \pi$ to $2 \pi$. In general, DG with ${\cal O}(x^p)$ basis functions corresponds to $p+1$ nodes per DG cell and includes modes up to $k \Delta x = (p+1) \pi$. That is, modes have up to $p+1$ extrema in oscillations per cell, though the eigenvalues of the shortest wavelength modes may not be very accurate.

\subsection{Stability of the nonlinear model-Fokker-Planck operator}

The spatial discretization of the Vlasov-Dougherty~equation discussed in~\cite{Hakim2019} can use the same estimates for the SSP-RK3 $\Dt$ as those for the one-velocity-dimension \mFPO. Such limit of \eqr{\ref{eq:GkFP}} is
\begin{equation} \label{eq:gkFPO1V}
\pd{\Jac f}{t} = \Jac\gkcoll{f} = \pd{}{\vpar} \left[
\nu (\vpar - \upar) \Jac f + \nu \vt^2 \pd{\Jac f}{\vpar}\right].
\end{equation}
The first term looks like an advection term, so one might think that it gives an imaginary part to the eigenvalues. However, the eigenvalues of the full collision operator are not necessarily a simple sum of the separate eigenvalues of the diffusion and advection terms. We have already noted in section~\ref{sec:theory} that the combined drag and diffusion terms in the continuous \mFPO~have a set of eigenmodes that are all purely damped (real $\lambda < 0$).

For now we use a conservative estimate of the eigenvalues of the DG-discretized \mFPO~based on a sum of contributions from the advection and diffusions terms. We use $\Dvpar = \Dv_{\parallel,\mathrm{cell}}/(2 p +1)$ for the advection term and $\Dvpar = \Dv_{\parallel,\mathrm{cell}}/(p +1)$ for the diffusion term. Assuming a constant grid spacing, the estimated maximum eigenvalue of the \mFPO~is
\begin{equation} \label{eq:ch5-lambdaMax-1x1v}
\left|\eigmax\right| = 2\nu C_{\adv,p} \max(|\vpar - \upar|) \frac{(2 p + 1)}{\Dv_{\parallel,\mathrm{cell}}}
+ 4\nu C_{\dif,p} \vt^2 \left(\frac{p + 1}{\Dv_{\parallel,\mathrm{cell}}}\right)^2.
\end{equation}
This can be generalized to higher dimensions without difficulty. For the 1X2V (one configuration-space dimension, and two velocity-space dimensions) \mFPO~we use the following maximum eigenvalue estimate
%\begin{eqnal} \label{eq:ch5-lambdaMax-1x2v}
%\left|\txtS{\eig}{\mFPO}\right| & = 2\nu C_{\adv,p} \max(|\vpar - \upar|) \frac{(2 p + 1)}{\Dv_{\parallel,\mathrm{cell}}}
%+ 4\nu C_{\dif,p} \vt^2 \left(\frac{p + 1}{\Dv_{\parallel,\mathrm{cell}}}\right)^2 \\
%&\quad+ 4\nu C_{\adv,p} \mumax \frac{(2 p + 1)}{\txtS{\Dmu}{cell}}
%+ 8\nu C_{\dif,p} \frac{m \vt^2}{B} \mumax \left(\frac{p + 1}{\txtS{\Dmu}{cell}}\right)^2.
%\end{eqnal}
\begin{equation} \label{eq:ch5-lambdaMax-1x2v}
\begin{aligned}
\left|\eigmax\right| = &2\nu C_{\adv,p}(2 p + 1)\left[\frac{\max(|\vpar - \upar|)}{\Dv_{\parallel,\mathrm{cell}}}+\frac{2\mumax}{\txtS{\Dmu}{cell}}\right]
+ 4\nu C_{\dif,p} \vt^2 \left(p + 1\right)^2\left[
\frac{1}{\left(\Dv_{\parallel,\mathrm{cell}}\right)^2} +\frac{m}{B} \frac{2\mumax}{\left(\txtS{\Dmu}{cell}\right)^2}\right].
\end{aligned}
\end{equation}
This eigenvalue is computed every stage of the SSP-RK3 and used to calculate the time step according to $\Dt|\txtS{\eig}{\mFPO}|<\cfl$. The $\cfl$ number is close to unity, but in \gkeyll~it can be modified as a user input.

\begin{figure*}[h]
    \centering
    \includegraphics[width=0.9\textwidth]{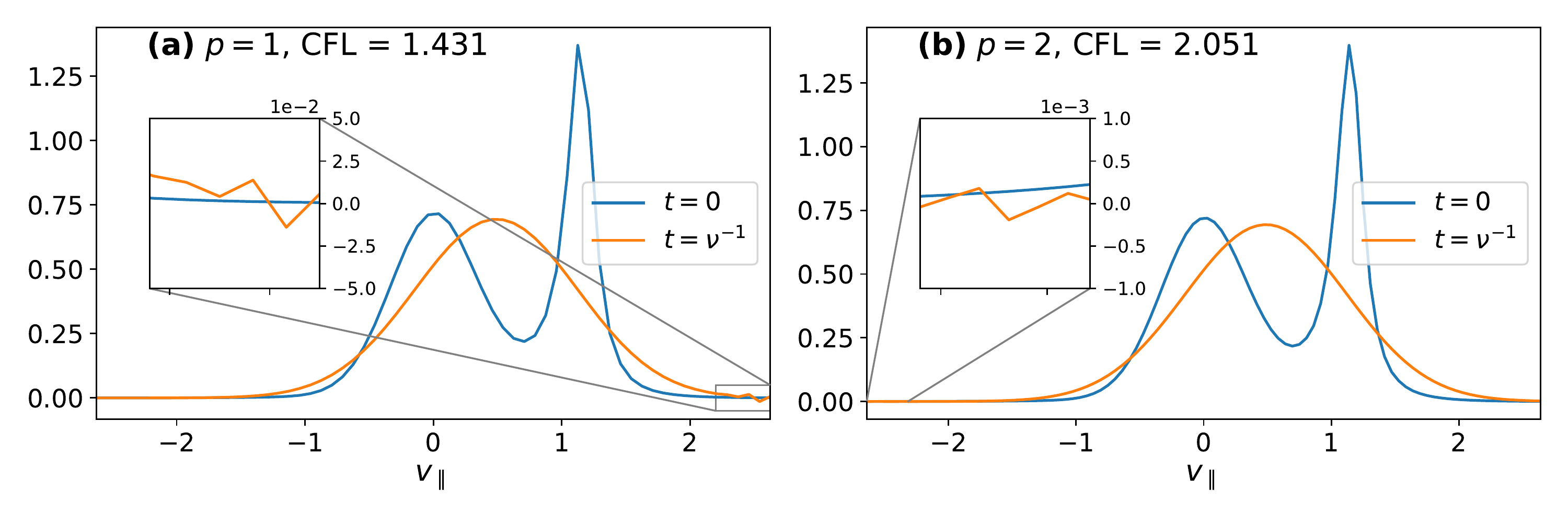}
    \caption{Relaxation of a 1X1V bump-on-tail distribution with (a) $p=1$ and (b) $p=2$ using the minimum \cfl~at which an instability is found. Unstable oscillations are seen at the edges of the domain, which grow quickly if the simulations are run for longer times.}
    \label{fig:1x1v-cfl}
\end{figure*}

In order to illustrate the impact of these choices consider the spatially homogeneous relaxation problem in 1X1V posed by \eqr{\ref{eq:gkFPO1V}} with a bump-on-tail distribution function for its initial condition:
\begin{equation} \label{eq:ch5-bot-lg}
f(t=0) = f_M(n,\upar,\vt) + f_{M}(n,u_{\parallel,b},v_{t,b})\frac{a_b^2}{(\vpar - u_{\parallel,b})^2 + s_b^2}.
\end{equation}
Here $f_M(n,\upar,\vt)$ is the one velocity-space dimension Maxwellian
\begin{equation}
f_M(n,\upar,\vt) = \frac{n}{\sqrt{2\pi\vt^2}}\exp\left[-\frac{\left(\vpar-\upar\right)^2}{2\vt^2}\right],
\end{equation}
and we employed the parameters $n=1$, $\upar=0$, $\vt=1/3$, $a_b=\sqrt{0.1}$, $u_{\parallel,b}=6\vt/\sqrt{3}$, $v_{t,b}=1.0$ and $s_b=0.12$. This distribution is discretized in a $[0,1]\times[-8\vt,8\vt]$ domain using $2\times32$ cells and a pieceswise linear basis ($p=1$), or $2\times16$ and a piecewise quadratic basis ($p=2$). Using a collisionality of $\nu=0.01$ we show that by time $t= \nu^{-1}$ the \mFPO~relaxes this initial condition to be close to a Maxwellian (\figr{\ref{fig:1x1v-cfl}}). 
By gradually increasing the CFL for each test, we discovered that for piecewise linear basis functions ($p=1$) the simulation begins to become unstable for $\cfl\gtrsim 1.431$, which gives $\Dt\approx0.075726$. Oscillations are observed close to or below zero in regions where $f$ is small (inset of \figra{\ref{fig:1x1v-cfl}}{a}), and the simulation diverges at later time. Piecewise quadratic basis functions ($p=2$) allowed $\cfl\lesssim2.051$, corresponding to $\Dt\approx0.062479$ (\figr{\ref{fig:1x1v-cfl}}).
The fact that these two cases ($p=1, 2$) were not stable all the way up to ${\rm CFL} = 
\gamma_{\rm max} \Delta t < 2.512$ (the stability limit for RK3 for damped modes) indicates that there are some inaccuracies in the approximations that led to Eq.~\ref{eq:ch5-lambdaMax-1x2v}, such as in treating advection and diffusion separately or in neglecting boundary conditions.  Nevertheless, it captures the main scaling of the allowable time step with the parameters of the problem.

\begin{figure*}[h]
    \centering
    \includegraphics[width=0.9\textwidth]{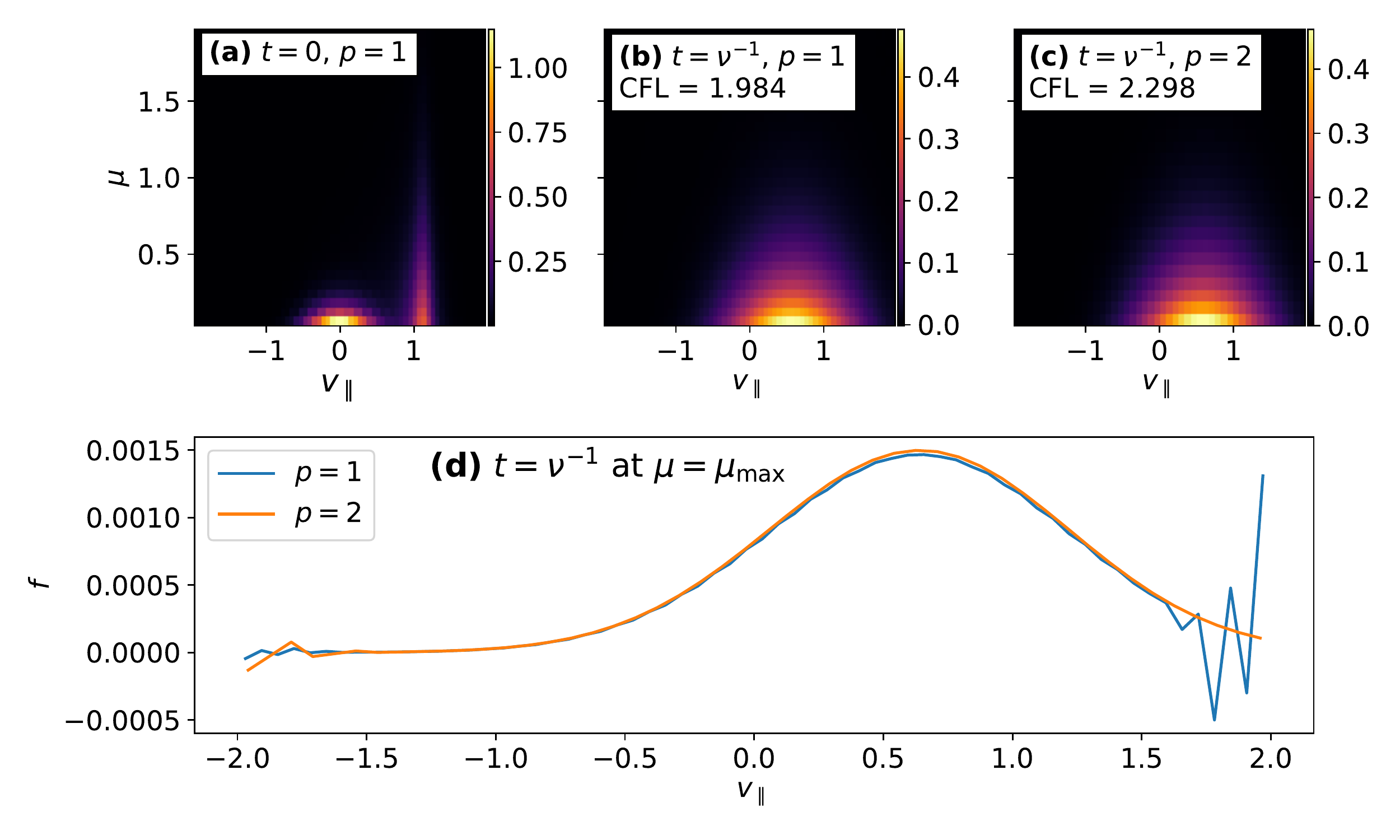}
    \caption{(a) Initial 1X2V bump-on-tail distribution. \mFPO~relaxation of such initial condition with (b) $p=1$ and (c) $p=2$, using the minimum unstable \cfl. (d) Demonstration of the instability beginning to form near the right (left) boundary for $p=1$ ($p=2$) at a slice through $\mu=\mumax$ and $t=\nu^{-1}$ when the minimum unstable CFL is used.}
    \label{fig:1x2v-cfl}
\end{figure*}

Similar oscillations are observed in 1X2V simulations. We projected \eqr{\ref{eq:ch5-bot-lg}} onto the 1X2V DG basis, using the Maxwellian $f_M(n,\upar,\vt)$ defined in \eqr{\ref{eq:maxwellian}}, with parameters $\vt=1/\sqrt{12}$, $v_{t,b}=1/\sqrt{2}$, and $u_b=4v_t$. The bump-on-tail initial condition is relaxed to a Maxwellian within one collisional period (compare figures~(\ref{fig:1x2v-cfl}a) with figures(\ref{fig:1x2v-cfl}b) and (\ref{fig:1x2v-cfl}c)). The domain in these plots is $[0,1]\times[-\vparmax,\vparmax]\times[0,m\vparmax^2/(2B)]$ with $\vparmax=6.93\vt$, $m=1$, $B=1$ and using $2\times32\times16$ cells for $p=1$, or $2\times16\times8$ cells for $p=2$. As \cfl~increases, an instability begins to develop at $\cfl=1.984$ for $p=1$ ($\Dt\approx0.063075$), and $\cfl=2.298$ for $p=2$ ($\Dt\approx0.075544$). It is difficult to discern in figures~(\ref{fig:1x2v-cfl}b-c), but \figra{\ref{fig:1x2v-cfl}}{d} shows a slice at $f(x=0.5,\vpar,\mu=\mumax)$ exhibiting oscillations near the boundary with regions of $f<0$.

\begin{figure}[h]
    \centering
    \includegraphics[width=0.48\textwidth]{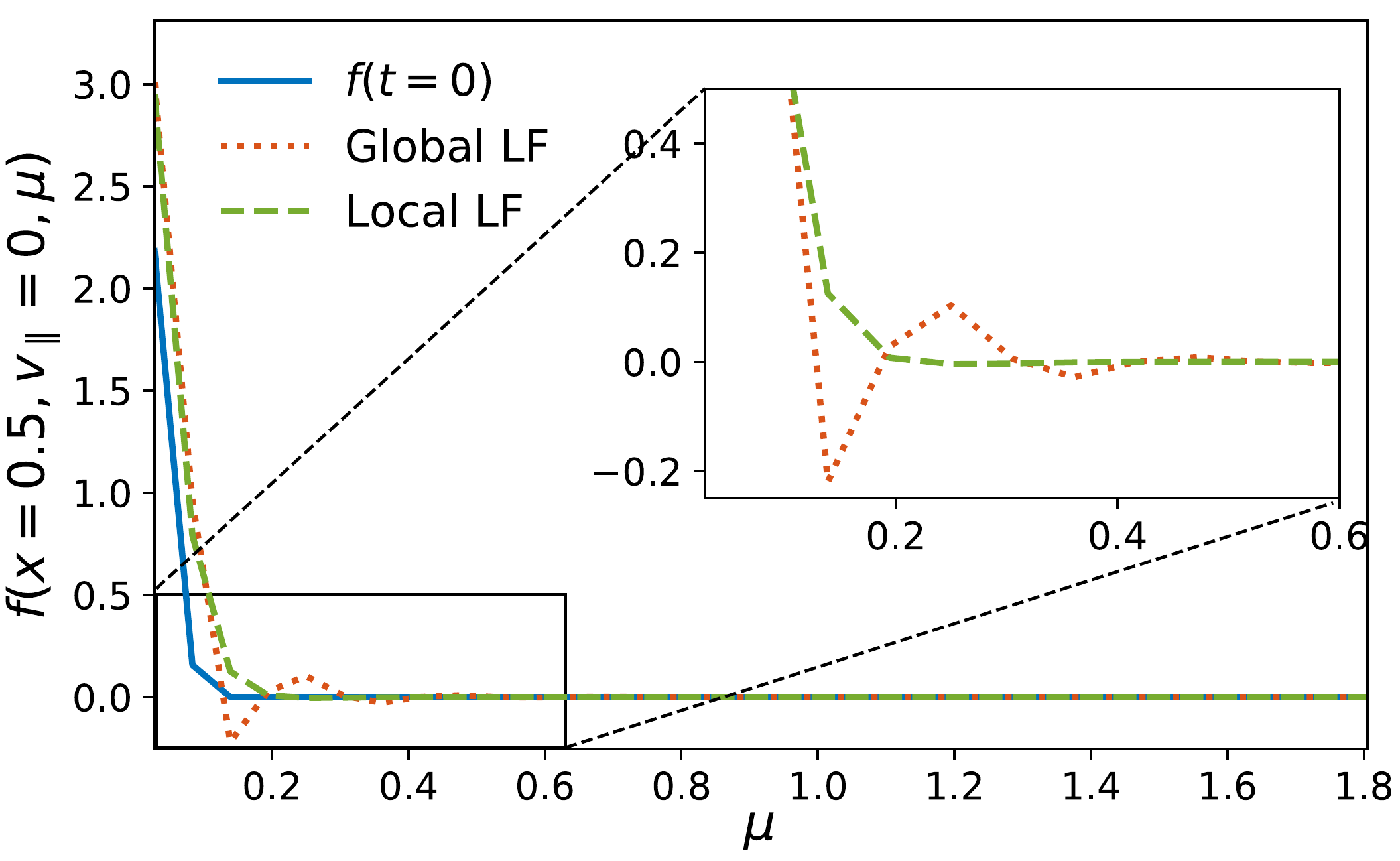}
    \caption[Relax square.]{Relaxation of a 1X2V rectangular distribution with a coarse $\mu$-grid with $p=1$. Here we show the initial (solid blue) and final (dotted orange and dashed green) distribution functions after one collisional period. Dotted orange used global Lax-Friedrichs fluxes, dashed green used LF fluxes based on the local values at quadrature points.}
    \label{fig:squareRelax}
\end{figure}

The errors illustrated above are eliminated by more conservative \cfl~choices, yet another area of concern is positivity of the distribution function. Note that negative $f$ values do not always immediately give way to numerical instabilities. An example of this resilience is illustrated by relaxing the rectangular distribution
\begin{equation} \label{eq:rectangularF}
f(t{=}0,x,\vpar,\mu) = \begin{cases}
1/(2\pi v_0^3) \qquad |\vpar|\leq v_0~\mathrm{and}~\mu\leq mv_0^2/(2B) \\
0 \qquad |\vpar| > v_0~\mathrm{or}~\mu>mv_0^2/(2B),
\end{cases}
\end{equation}
where $v_0=1/\sqrt{6}$, $m=1$ and $B=1$, on a coarse mesh of $4\times32^2$ cells\footnote{This system has no configuration-space variation so we could have used a $1\times32^2$ grid instead.} in the domain $[0,1]\times[-\vparmax,\vparmax]\times[0,m\vparmax^2/(2B)]$, using $\vparmax=16v_0^2$. With these parameters the $\mu$ direction is under-resolved, such that after a collisional period the rapid drop in $f$ at small $\mu$ is hard to capture with a piecewise linear basis. The result, as shown by the dotted orange line in \figr{\ref{fig:squareRelax}}, is that the solution overshoots near zero and causes regions of $f<0$. This can be problematic as it can lead to unphysical, local negative densities and temperatures. However, on this occasion the code remains stable and the relaxation proceeds without a problem; this is the maximum entropy solution. In section~\ref{sec:weakRDG} we commented on how RDG for the diffusion operator can lead to regions of $f<0$, and unfortunately lower order recovery polynomials did not provide a satisfactory solution (\figra{\ref{fig:recoverySample}}{b}). Positivity is better respected by using local Lax-Friedrichs (LF) fluxes: instead of the fluxes in \eqr{\ref{eq:flux}} with the maximum evaluated over the global domain (global LF fluxes), we can use
\begin{equation} \label{eq:localFlux}
\begin{aligned}
G_{\vpar}(f_L,f_R) &= (\vpar-\upar)\left\lbrace\frac{1}{2}(\Jac f_R+\Jac f_L) - \frac{\mathrm{sgn}\left[(\vpar-\upar)\hat{\vpar}\cdot\hat{n}\right]}{2}(\Jac f_L-\Jac f_R)\right\rbrace+ \vt^2 \pd{\Jac \frec}{\vpar}, \\
G_\mu(f_L,f_R) &= 2\mu\left[\frac{1}{2}(\Jac f_R+\Jac f_L) - \frac{\mathrm{sgn}\left(\hat{\mu}\cdot\hat{n}\right)}{2}(\Jac f_L-\Jac f_R)\right] + \frac{2 m \vt^2}{B} \mu \pd{\Jac \frec}{\mu},
\end{aligned}
\end{equation}
where $\hat{n}$ is the unit vector pointing out of the corresponding surface where the numerical flux is evaluated. Since the up-winding is based on local values of the phase-space velocity we refer to this as the local LF fluxes, and it yields a steady state that avoids the negative incursion of the global LF fluxes (\figr{\ref{fig:squareRelax}}).

\subsection{Stability condition in the full gyrokinetic-\mFPO~system}
The (long wavelength) gyrokinetic system in \eqsr{\ref{eq:gk-boltzmann}}{\ref{eq:5-poisson}} is also limited by the CFL constraints of the collisionless, or Hamiltonian, terms. We can estimate this condition by considering the kinetic equation
\begin{equation}
\pd{f}{t} + \nabla{_{\v{z}}\cdot (\gv{\alpha}f}) = 0,
\end{equation}
where the phase-space gradient $\nabla_{\v{z}} \equiv \left(\nabla,\frac{\partial}{\partial v_\parallel}\right)$ acts on the flux $\gv{\alpha}f$ with a phase-space velocity $\gv{\alpha} \equiv (\dot{\v{R}},\dot{v}_\parallel)$. This is a nonlinear advection equation, for which we can use \eqr{\ref{eq:advec-cell}} to estimate the stability condition on each cell:
\begin{equation}
\eig_{\mathcal{H}} = 2 \ C_{\adv,p} \, (2 p +1) \sum_k \frac{\max(0, \gv{\alpha} \cdot \hat{n}_k)}{\Delta z_k},
\end{equation}
where the $k$ sum is over all faces of the cell, $\hat{n}_k$ is the outward normal of the $k^{\rm{th}}$ face, and $\Delta z_k$ is the grid spacing in the direction corresponding to the $k^{\rm{th}}$ face. The form of the maximum function guarantees that the sum is only over faces where there is an outgoing flux. Although the eigenvalues of the full collisional gyrokinetic equation are not a sum of the collisionless eigenvalues and the \mFPO\ eigenvalues, we follow this conservative approach and compute the time step according to
\begin{equation}
\Dt\left(\eig_{\mathcal{H}}+\txtS{\eig}{\mFPO}\right) < \cfl.
\end{equation}

An example of what establishes $\eig_{\mathcal{H}}$ in the electrostatic limit is the electrostatic shear Alfv\'en or $\omega_H$ mode \cite{Belli2005,Shi2017thesis}. A dispersion relation for this electrostatic instability can be derived by linearizing the collisionless form of \eqr{\ref{eq:gk-boltzmann}} and \eqr{\ref{eq:5-poisson}}. In the long-wavelength limit this becomes
\begin{equation}
\omega_H = \sqrt{\frac{n_e}{n_0}}\frac{|\kpar v_{te}|}{|\kperp \rhos|} \label{eq:omegaH},
\end{equation}
where $n_0$ is the linear ion polarization density used in the Poisson equation. We seek an estimate for $\cfl$ so that $\omega_{H,{\rm max}}\Dt < 1.73$, which is the stability limit for the RK3 time-stepping method. To estimate $\omega_{H,{\rm max}}$, assume $k_{\parallel,{\rm max}} \approx \Dz = (2p+1)/\txtS{\Dz}{cell}$ and $k_{\perp,{\rm min}} = \pi /\Lx$, where $\Dz$ is the cell spacing in $z$ and $\Lx$ is the domain width in $x$. The initial time-step is set by the fastest parallel electron transit rate, $\vparmax$. The corresponding eigenvalue is $\vparmax/\Dz = (2p+1)\vparmax/\txtS{\Dz}{cell}$, giving a time-step estimate of
\begin{equation}
\Delta t = \frac{\txtS{\Dz}{cell}\thinspace \mathrm{\cfl}}{(2p + 1)\vparmax}. 
\end{equation}
Combining with \eqr{\ref{eq:omegaH}} and the RK3 stability limit gives
\begin{equation}
    \omega_{H,\max} \Dt = {\rm \cfl} \sqrt{\frac{n_{e,\max}}{n_0}} \frac{v_{te} \, \Lx}{\pi v_{\parallel e,\max} \, \rhos} < 1.73, \label{eq:omegaH-max}
\end{equation}
which we use to set an appropriate value of \cfl~as an input parameter prior to run time.
In the future, we plan to calculate $\omega_{H,\max}$ within the code to dynamically set the time-step limit due to the electrostatic shear Alv\'en mode.

% {\color{blue}
% \emph{Notes and comments}
% \begin{align}
%     &\omega_a \equiv c/\Delta x \\
%     &\omega_D \equiv \alpha / (\Delta x)^2 \\
%     &\Delta t \max \sum_k \omega_k = \rm{CFL}
% \end{align}
% \begin{align}
%     \sqrt{\omega_1^2 + \omega_2^2} \le |\omega_1| + |\omega_2|
% \end{align}
% \begin{itemize}
%     \item Formulate everything in terms of $\omega$ (frequency). 
%     \item Gyroaveraging on collision operator?? Paper? CQL3D code: bounce-averaged.
% \end{itemize}
% }

%% file: results.tex
\section{Benchmark problems} \label{sec:results}
In this section, we present tests designed to further understand the discrete scheme and to verify the accuracy of the gyrokinetic-\mFPO~system. Relaxation tests without the collisionless terms demonstrate properties of the discrete \mFPO~such as conservation, entropy and positivity. We also explore collisional Landau-damping to understand the physical implications of this model and compare it to analytic theory. Finally, simulations of 5D turbulence on helical, open field lines with collisions modeled by the \mFPO~are presented.

\subsection{Relaxation tests of the \mFPO} \label{sec:relax}
\begin{figure}[h]
  \centering
  \includegraphics[width=0.49\textwidth]{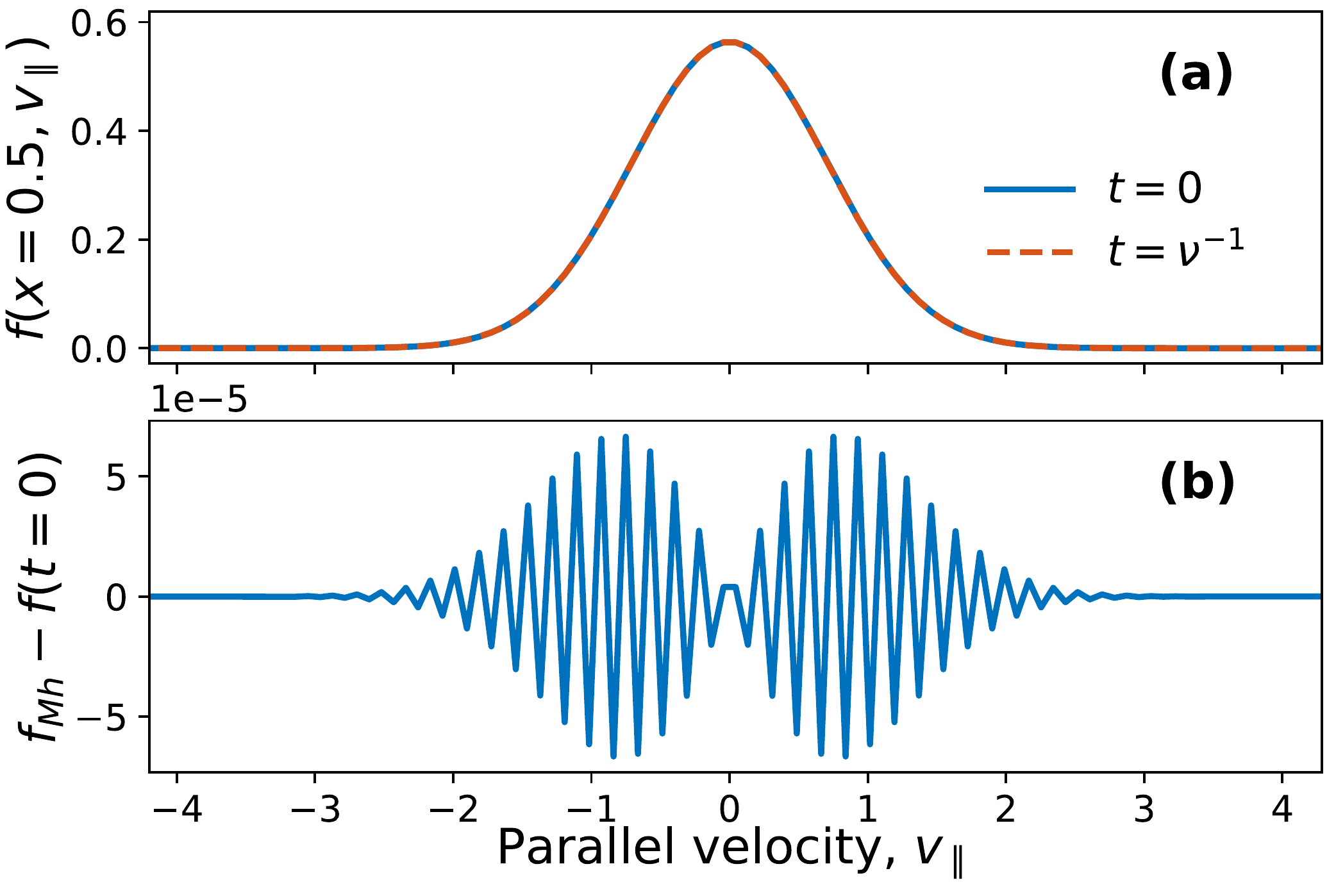}
  \includegraphics[width=0.49\textwidth]{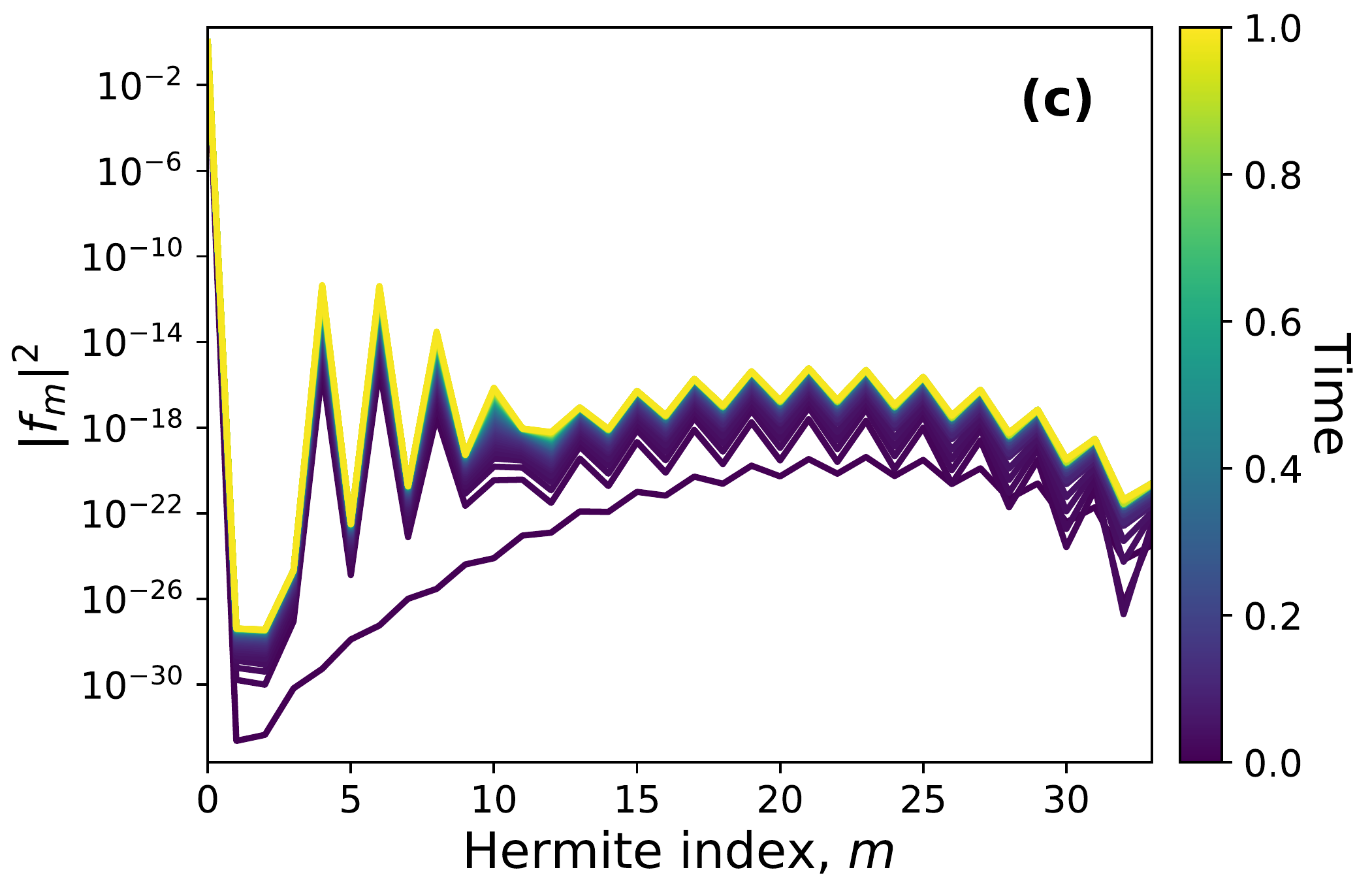}
  \caption[Evolution of Maxwellian.]{(a) An initial Maxwellian projected onto the DG basis and its final ($t=\nu^{-1}$) state after relaxation. (b) Difference between initial projection of the Maxwellian and final discrete equilibrium, i.e. $f_{Mh}=f(t=\nu^{-1})$. (c) Time evolution of the Hermite spectrum when initial state is a projected Maxwellian.}
  \label{fig:maxwellianEv}
\end{figure}

When an initial distribution function is subjected to the \mFPO~alone, without the Hamiltonian terms, it will relax to the maximum entropy solution. In the continuous sense, the maximum entropy solution is the Maxwellian in \eqr{\ref{eq:maxwellian}}, but the discrete equilibrium solution, $f_{Mh}$, is not necessarily the projection of \eqr{\ref{eq:maxwellian}} onto the DG basis. In principle, $f_{Mh}$ could be derived by repeating the derivation of \eqr{\ref{eq:maxwellian}} but assuming the discrete form of the \mFPO~and a finite velocity domain. This implies that if we project $f_M$ onto the basis (e.g. using Gaussian quadrature) and use that as an initial condition, the system will not be static and will evolve some. Figure (\ref{fig:maxwellianEv}a), for example, shows this initial, projected Maxwellian and its final state after one collisional period. At first sight they are indistinguishable, but the difference, shown in \figra{\ref{fig:maxwellianEv}}{b}, shows that the projected Maxwellian was not in the kernel of the discrete $\gkcoll{f}$. These tests were carried out in a $[0,1]\times[-12\vt,12\vt]$ domain with $2\times96$ cells using a zero-drift Maxwellian with $\vt=1/\sqrt{2}$, $\nu=0.01$ and piecewise linear bases.

A Maxwellian is equivalent to the (properly normalized) Gaussian-weighted zeroth Hermite basis~\cite{Adkins2018}. This is also an eigenfunction of $\gkcoll{f}$, as we will show in section~\ref{sec:landau}. Therefore, in a continuous infinite velocity-space, its Hermite spectrum should remain a Dirac delta function peaked at the Hermite index $m=0$ as it is evolved in time according to \eqr{\ref{eq:gkFPO1V}}. Ideally this ought to be reflected in the spectral analysis of the discrete data, but spectral transforms of DG data are subtle. If one transforms the data interpolated onto a finer grid (e.g. $\Dx=\Dxcell/(p+1)$ as in section~\ref{sec:stability}) errors would be introduced in the higher modes due to the underlying piecewise discontinuous character. The appropriate way to transform DG data is by formulating it as a weak equivalence. Therefore the spectral transform of one-dimensional data onto a normalized Gaussian-weighted Hermite basis is given by
\begin{equation} \label{eq:weakHermite}
    \txtS{f}{DG}(x) = \sum_{k=1}^{\Np} f_k\psi_k(x) \doteq \txtS{f}{Hermite}(x) = \sum_{m=0}^{\mmax}f_m\frac{1}{\sqrt{2^m m!}}\hermit{m}(x)\frac{e^{-x^2}}{\sqrt{\pi}},
\end{equation}
where $\hermit{m}(x)$ is the $m$-th physicists' Hermite polynomial. After projecting each of the Gaussian-weighted Hermite basis functions onto the DG basis using Gaussian quadrature, the weak equality in \eqr{\ref{eq:weakHermite}} yields a linear system of equations in the $f_m$ unknowns. When the number of DG degrees of freedom is larger than $\mmax+1$ this linear problem is solved via least squares. In multiple dimensions, one can project the function onto the the basis that excludes the to-be-transformed dimension, and perform a series of 1D transforms. Such operation was carried out with the projected Maxwellian subjected to \eqr{\ref{eq:gkFPO1V}}, yielding the spectrum in \figra{\ref{fig:maxwellianEv}}{c}. This analysis suggests there exist contributions from modes other than $m=0$ at $t=0$ (darkest purple), albeit orders of magnitude smaller. As time proceeds all $m\neq0$ modes grow, and high $m$ modes saturated at $\left|f_m\right|^2{\sim}10^{-18}$ form part the discrete maximum entropy solution.

The deviation of the spectrum in \figra{\ref{fig:maxwellianEv}}{c} from the continuous, infinite space equivalent may be the result of a number of factors. The Gaussian-weighted Hermites are orthonormal in the infinite velocity space, but truncating the domain introduces errors in the orthonormality relation (i.e. the orthonormality integral no longer yields a Kronecker delta function). These errors should be small in the $[-12\vt,12\vt]$ spaced used above. A greater difficulty in performing a Hermite spectral analysis arises from the discrete representation of the solution, and the Gaussian-weighted Hermites, in terms of discontinuous polynomial basis. The departure from orthonormality of the discrete Gaussian-weighted Hermites may be more significant than that of the continuous ones on a restricted domain. There is also some ambiguity as to how to project the Gaussian-weighted Hermites onto the DG basis; we used Gaussian quadrature, but one could also evaluate them at cell nodes to produce a continuous representation or use an exact projection. Additionally the least-squares solution of \eqr{\ref{eq:weakHermite}} may also introduces other errors, particularly if the corresponding matrix is ill-conditioned. Furthermore, notice that in \figra{\ref{fig:maxwellianEv}}{c} the spectrum is truncated at $\mmax+1=34$. Initially we expected $\mmax+1$ to correspond to the number of modes whose roots are contained within our domain (44 in this case), but instead we found empirically that if $\mmax+1>34$ the Hermite analysis yields high $m$ mode amplitudes orders of magnitude larger. Further exploration of spectral transforms of DG data, including the a priori determination of $\mmax$, is an interesting enterprise currently ongoing in our group but which unfortunately is beyond the scope of this publication.

\begin{figure}[h]
  \centering
  \includegraphics[width=0.49\textwidth]{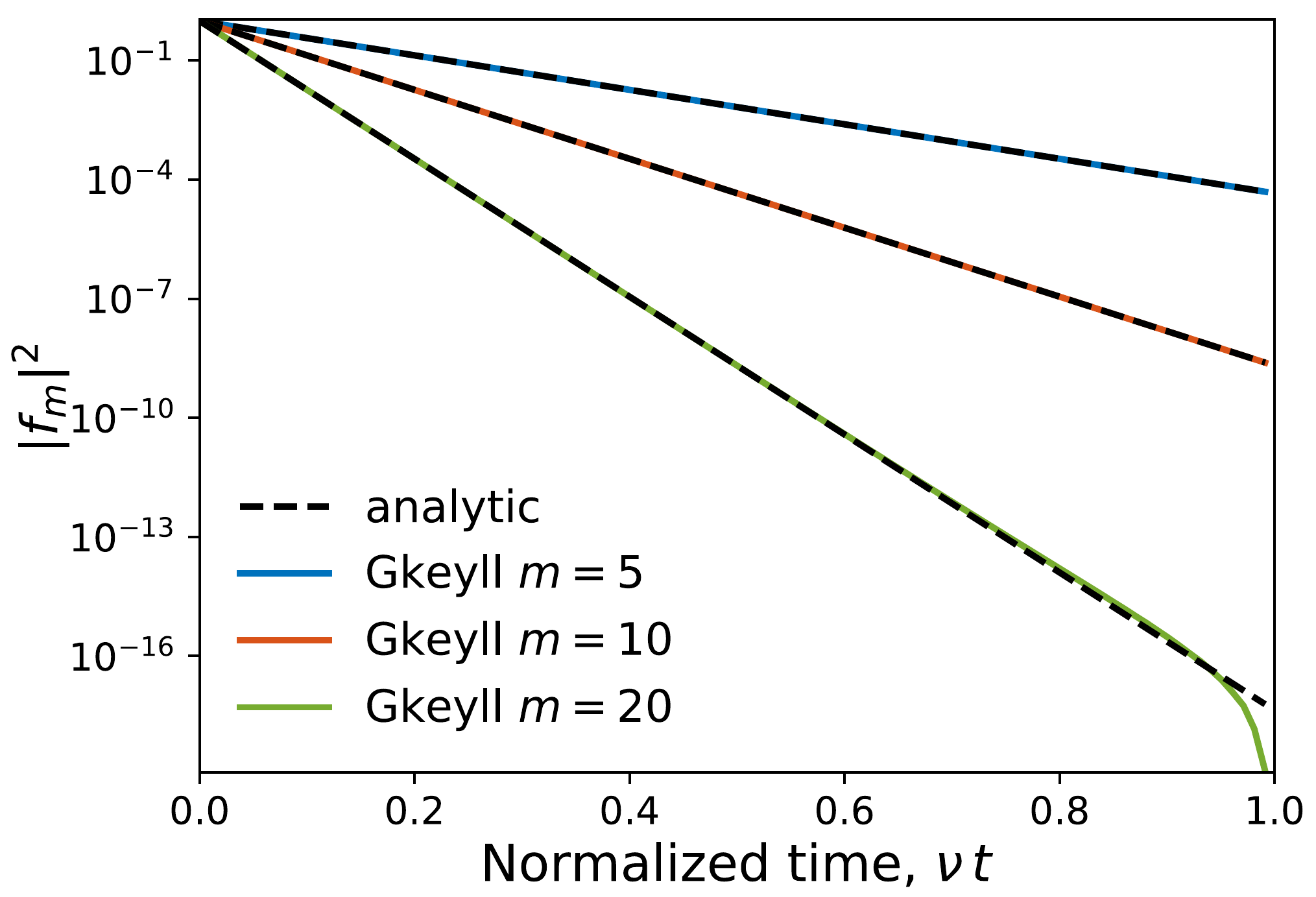}
  \caption[Time evolution of Hermite coefficients.]{Time evolution of the (squared) Hermite expansion coefficients of the distribution function.}
%  To distinguish a damped mode from zero we need to ensure that $e^{-2m \nu t}$ is larger than machine precision. Hence for $m=20$ machine precision errors will dominate when $\exp(-40\nu t)<10^{-16}$ or when $\nu t > 0.92$}
  \label{fig:hermite}
\end{figure}

As we will show in section~\ref{sec:landau}, the Hermite basis diagonalizes the \mFPO, turning \eqr{\ref{eq:gkFPO1V}} into
\begin{equation}
    \pd{f_m}{t} = -\nu m f_m.
\end{equation}
We can test the analytic solution to this equation, $f_m(t) = f_m(t=0)e^{-\nu m t}$, numerically by using $f_{m=0}=f_{m=5}=f_{m=10}=f_{m=20}=1$ and zero for all other modes, rather than only initializing $f_{m=0}=1$, as in \figr{\ref{fig:maxwellianEv}}. The time evolution of the three higher modes is shown in \figr{\ref{fig:hermite}}. Its agreement with analytic theory is very good in the $t\in[0,\nu^{-1}]$ time window, and if the discrete Hermite basis functions were eigenfunctions of our discrete \mFPO, the three higher modes would simply decay exponentially indefinitely. However, when the amplitude of the $m=20$ mode reaches the noise introduced by the evolution of $m=0$, the spectral analysis of $m=20$ begins to deviate from the analytic result. This noise we showed in \figra{\ref{fig:maxwellianEv}}{c} is at a $\left|f_m\right|^2{\sim}10^{-18}$ level, and it is at that point that the green $m=20$ line in \figr{\ref{fig:hermite}} deviates from its analytic expectation. Were this spectral analysis to be carried out with $\mmax+1<33$ the error in $m=20$ would be slightly larger at $\nu t>0.6$, but solving the least-squares problem in \eqr{\ref{eq:weakHermite}} with $\mmax+1>34$ yields errors orders of magnitude larger.

%MF:This sentence here referred to the analysis with m_max=44:
%The green line in \figr{\ref{fig:hermite}}~corresponding to $m=20$ appears to increase in the large $\nu t$ limit, but it in fact asymptotes to a flat line. The level at which this line settles is seen to decay upon grid refinement. %with fourth-order accuracy?

\begin{figure}[h]
    \centering
    \includegraphics[width=\textwidth]{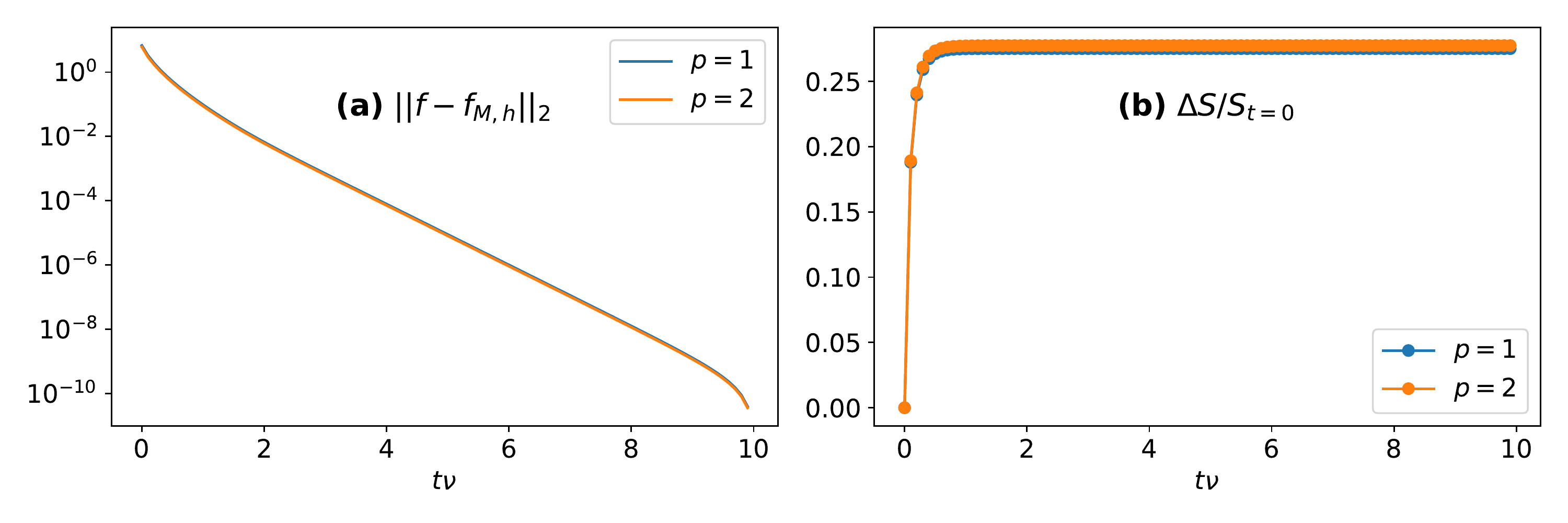}
    \caption[1X2V relaxation tests of the GK-FPO using bump-on-tail Maxwellian initial conditions showing increasing entropy.]{As a 1X2V a bump-on-tail distribution relaxes, the norm of the difference of $f$ and the discrete equilibrium solution $f_{Mh}=f(t=10\nu^{-1})$ decreases (a). The entropy $S$ increases monotonically, and so does the relative difference in $S$ (b).}
    \label{fig:5-relax-bump}
\end{figure}

As the solution relaxes onto the discrete maximum entropy solution, $f_{Mh}$, it also exhibits a physical non-decreasing entropy. We again project the bump-on-tail distribution of \eqr{\ref{eq:ch5-bot-lg}} onto the in 1X2V $(x,\vpar,\mu)$ DG basis, use a stable $\cfl=1$, and run to $\nu t=10$. The norm of the difference between $f(t)$ and the maximum entropy solution, $f_{Mh}=f(t=10\nu^{-1})$, decreases rapidly as shown in \figra{\ref{fig:5-relax-bump}}{a}. Meanwhile, the entropy, $S(t)=-\int f(t)\ln f(t)\thinspace\dx\thinspace\dThvv$, increases monotonically. The relative difference between initial entropy and $S(t)$ is given in \figra{\ref{fig:5-relax-bump}}{b}. Although we have not yet proven an $H$-theorem for the discrete operator (or proved its self-adjointness), the entropy is seen to increase in the cases we have explored. Part of the challenge in proving self-adjointness of the operator lies in guaranteeing that $f$ remains positive. Positivity of $f$ is something we are able to build into the discretization of the \mFPO's drag term (not presented here), but additional work is needed to ensure the diffusion term does not cause $f$ to go negative.

\begin{figure}[h]
    \centering
    \includegraphics[width=\textwidth]{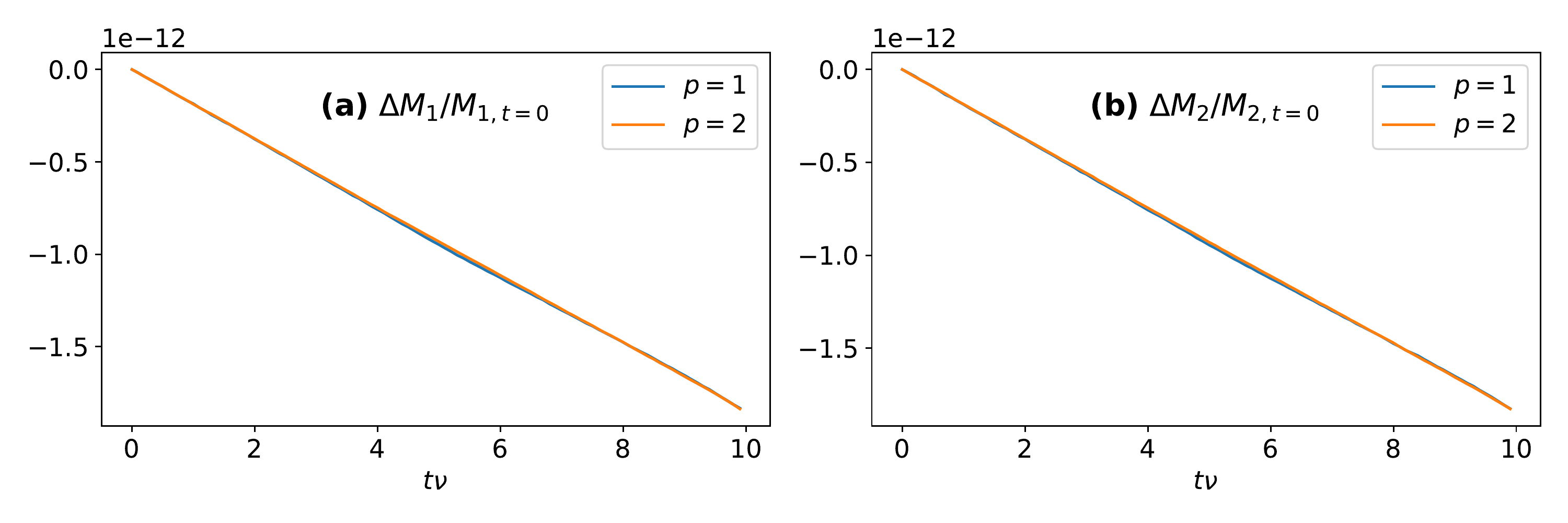}
    \caption[1X2V relaxation tests of the GK-FPO using bump-on-tail Maxwellian initial conditions showing conservation of moments.]{1X2V relaxation of a bump-on-tail distribution with $p=1$ and $p=2$. Relative norm of the difference in (a) momentum density $M_1$ and (b) energy density $M_2$, as a function of time. Both are conserved to machine precision.}
    \label{fig:5-mcons-bump}
\end{figure}

These 1X2V relaxation tests also confirm the conservative properties of our scheme and, although not shown here, conservation of particle number, momentum, and energy are also guaranteed in higher dimensions. For the case of the 1X2V bump-on-tail initial condition, \figr{\ref{fig:5-mcons-bump}} shows the norm of the relative difference in the momentum and energy densities, $M_1$ and $M_2$. Over ten collisional periods the relative change in these quantities remains within machine precision, consistent with sections~\ref{sec:discM1con}-\ref{sec:discM2conP1}. In this case machine precision accuracy refers to the fact that the relative error per time step in the momentum is ${\sim}2{\times}10^{-12}/1400\sim10^{-15}$,
%and the relative error per velocity degree-of-freedom is ${\sim}2{\times}10^{-12}/(1400\times32\times16)\sim10^{-18}$
where 1400 is the approximate number of time steps.
%and $32{\times}16$ is the resolution in velocity-space
The non-vanishing boundary contributions in the surface term of \eqr{\ref{eq:schemec}} and in the calculation of the primitive moments, $\upar$ and $\vt$, are necessary for exact conservation, even if $f$ is small at the boundaries. Neglecting these corrections gives errors in momentum and energy conservation that are orders of magnitude larger.

\subsection{Collisional Landau damping of ion acoustic waves} \label{sec:landau}

The study of collisionless (Landau) damping of plasma waves due to velocity-space resonance dates back to the origins of plasma physics, and its modification due to the presence of collisions remains an important area of research. Ion sound waves suffer from this decay, and scientists have been constructing a theory of such phenomenon for decades. Consider that neutral sound waves are undamped in its highly (molecular) collisionality environment, so one may expect that as collisions become more frequent Landau damping of ion acoustic waves would weaken. The consensus, however, is that the actual trend depends on whether one considers self-species collisions, multi-species collisions, or both. The description of collisional Landau damping can also vary with the collision operator employed. An early study with a Krook operator~\cite{Bhadra1964} noted that under ion-ion collisions alone the damping rate ($\gamma=-\mathrm{Im}
\thinspace\omega$) monotonically decreases towards the regular sound wave limit ($\gamma\to0$) as as $\nu_{ii}$ increases if the temperature ratio $\tau=T_{i}/T_{e}=1$. This was not limited to the simple Krook operator as numerical integration of the Vlasov-FPO equation also arrived at the same conclusion~\cite{Ono1975}. But these studies, and also~\cite{Randall1982}, quickly noticed that in non-equilibrium cases of nonequal temperatures, specifically $\tau<1$, the damping rate can first undergo a period of growth before starting to decrease towards the undamped fluid limit. Even at equal temperatures, including electron-ion collisions can increase the damping rate~\cite{Epperlein1992}.

Landau damping of plasma sound waves is central to ion-temperature gradient instabilities,  ion acoustic instabilities and other transport processes in astrophysical and laboratory plasmas. As a commonplace ingredient in plasmas, it is not only necessary to understand its collisional modifications with the full FPO, but also with the simple models frequently used by analytic and computational studies. The model-FPO Dougherty operator considered here has been explored little in the context of ion-acoustic waves. One of the few studies available~\cite{Ong1969} explored ion-acoustic instabilities in the presence of self-species and multi-species collisions, and was only able to do so at low collisionalities (i.e. $(\nuee+\nuie)/(\kpar\vte)\ll1$). The study of collisional Landau damping of ion-acoustic waves at arbitrary collisionality here then serves as both documentation of this process with the Dougherty operator, and also as validation of our scheme and implementation within \gkeyll.

Consider a system consisting of a single-ion hydrogen plasma in a curvature-free homogeneous magnetic field such that $\Jac=B=\bhat\cdot\vB=\zhat\cdot\vB$. The electrons will be assumed adiabatic and will not collide with the ions (only ion self-species collisions are included here), so we refer to the collisionality $\nuii=\nu$. Then one only needs to evolve the ion equation~\eqr{\ref{eq:gk-boltzmann}}, which upon linearization about an equilibrium, $f=f_0+f_1$ (we omit the ion subscript here for simplicity), simply becomes
\begin{equation}
\begin{aligned}
\pd{f_1}{t} + \vpar\pd{f_1}{z}+\frac{1}{B}\left[\phi,f_0\right] - \frac{e}{m}\pd{\phi}{z}\pd{f_0}{\vpar} &= C[f_0,f_1] \\
&= \nu\left\lbrace\pd{}{\vpar}\left[\left(\vpar-\uparZ\right)f_1-\uparO f_0+\vtZ^2\pd{f_1}{\vpar}+\vtO^2\pd{f_0}{\vpar}\right] \right. \\
&\left.\qquad+\pd{}{\mu}2\mu\left[ f_1+\frac{m}{B}\left(\vtZ^2\pd{f_1}{\mu}+\vtO^2\pd{f_0}{\mu}\right)\right]\right\rbrace.
\end{aligned}
\end{equation}
The simplified Poisson bracket $[F,G]=\bhat\cdot\grad{F}\times\grad{G}$ vanishes since $f_0$ is homogeneous in configuration space, and the first-order primitive moments are defined as
\begin{align}
\uparO &= \frac{2\pi B}{m n_0}\int \vpar f_1\thinspace\dvdmu, \\
\vtO^2 &= \frac{2\pi B}{3mn_0}\int \left(\frac{2\mu B}{m}+\vpar^2-3\vtZ^2\right) f_1\thinspace\dvdmu,
\end{align}
and $n_0$, $\uparZ$ and $\vtZ$ are the number density, mean velocity and thermal speed of $f_0$, respectively, although we have assumed $\uparZ=0$. It is convenient to write the perturbed distribution as $f_1=f_0\varphi$ with $\varphi\ll1$. The linearized collision operator then becomes
\begin{equation} \label{eq:linColl}
\begin{aligned}
C[f_0,f_1] = \nu f_0 &\left\lbrace-\left(\vpar-\uparZ\right)\pd{\varphi}{\vpar}
+\vtZ^2\pdd{\varphi}{\vpar}-2\mu\pd{\varphi}{\mu}+\frac{m\vtZ^2}{B}\pd{}{\mu}2\mu\pd{\varphi}{\mu} \right.\\
&\left.\quad-\frac{\vtO^2}{\vtZ^2}\left[3-\frac{2\mu B}{m\vtZ^2}-\frac{\left(\vpar-\uparZ\right)^2}{\vtZ^2}\right]+\uparO\frac{\vpar-\uparZ}{\vtZ^2}\right\rbrace.
\end{aligned}
\end{equation}
We have assumed there is no equilibrium component to the electrostatic potential ($\phi=\phi_1$), and from here on we will make use of the normalized variables $2\mu B/(m\vtZ^2)\to\mu$ and $\vpar/\vtZ\to\vpar$. Assuming wave-like modes according to the ansatz $f_1=\hat{f}\exp[i(kz-\omega t)]$, $\phi=\hat{\phi}\exp[i(kz-\omega t)]$, and employing the quasineutrality between adiabatic electrons and the equilibrium ion distribution (via Poisson's equation), renders our kinetic equation into
\begin{equation} \label{eq:linKinEq}
i\left(\vpar - \Omega\right)\hat{f} + i\pi\vtZ^3\vpar\frac{f_0}{n_0}\frac{T_{e0}}{T_{i0}}\int\hat{f}\thinspace\dvdmu -\eta f_0\chi\left(\varphi\right) = 0.
\end{equation}
We now refer to the normalized mode frequency, $\Omega=\omega/(\kpar\vtZ)$, and the normalized collisionality $\eta=\nu/(\kpar\vtZ)$, and $\chi(\varphi)$ is the term between curly brackets in \eqr{\ref{eq:linColl}}.

One can proceed by expanding in a set of Hermite-Laguerre polynomials~\cite{Anderson2007} as
\begin{equation}
\varphi = \sum_{m,n=0}^\infty a_{mn}\varphi_{mn} = \sum_{m,n=0}^\infty a_{mn}\frac{1}{\sqrt{m!}}\mathrm{He}_m\left(\vpar\right)L_n\left(\mu/2\right),
\end{equation}
which satisfy the orthogonality relation
\begin{equation}
\inProd{\varphi_{n n'}}{\varphi_{m m'}} = \frac{1}{2\sqrt{2\pi}}\int\varphi_{n n'}\varphi_{m m'} e^{-(\vpar^2+\mu)/2}\thinspace\dvdmu = \kron{n}{m}\kron{n'}{m'}.
\end{equation}
Together with the recursion relations
\begin{equation}
\begin{aligned}
\mathrm{He}_{m+1}(\vpar) &= \vpar\mathrm{He}_m(\vpar)-m\mathrm{He}_{m-1}(\vpar), \\
(\mu/2)L_n'(\mu/2) &= nL_n(\mu/2)-nL_{n-1}(\mu/2)
\end{aligned}
\end{equation}
one may find the projection of the transformed linear kinetic equation~\ref{eq:linKinEq} onto the basis $\varphi_{mn}$. This projection, after some algebra, is
\begin{equation} \label{eq:linKinEqProj}
\begin{aligned}
\inProd{\varphi_{mn}}{\mathrm{Eq.~\ref{eq:linKinEq}}} = \begin{cases}
\Omega a_{10}-a_{00}-\sqrt{2}a_{20}- \frac{T_{e0}}{T_{i0}}a_{00} = 0 \qquad (m,n)=(1,0) \\
\Omega a_{20}- \sqrt{2}a_{10}-\sqrt{3}a_{30} + i\eta\left(\frac{4}{3}a_{20}+\frac{2\sqrt{2}}{3}a_{01}\right) = 0  \qquad (m,n)=(2,0)\\
\Omega a_{01}-a_{11}+ i\eta\left(\frac{2}{3}a_{01}+\frac{2\sqrt{2}}{3}a_{20}\right) = 0  \qquad (m,n)=(0,1)\\
\left[\Omega+i\eta(m+2n)\right] a_{mn}- \sqrt{m}a_{(m-1)n}-\sqrt{m+1}a_{(m+1)n} = 0 \qquad \mathrm{all~other}~(m,n).
\end{cases}
\end{aligned}
\end{equation}
From the last of these equations one can show that for a physically realizable solution $a_{(m+1)n}/a_{mn}\to a_{mn}/i\eta\sqrt{m}$ as $m\to\infty$~\cite{Ng1999}. The presence of collisions limits the extent of the spectrum in $m$, allowing us to truncate the expansion at an upper limit $\mmax$. One can then use
\begin{equation} \label{eq:mmaxEq}
\left[\Omega+i\eta\left(\mmax+2n\right)\right] a_{\mmax n} - \sqrt{\mmax}a_{(\mmax-1)n} = 0
\end{equation}
in conjunction with the last relation in \eqr{\ref{eq:linKinEqProj}}
to iterate backwards from $\mmax$ and find:
\begin{equation} \label{eq:cFrac1}
a_{m0} = \cfrac{\sqrt{m}}{\Omega+i\eta m-\cfrac{m+1}{\Omega+i\eta\left(m+1\right)-\cfrac{m+2}{\Omega+i\eta(m+2)-\dots \cfrac{\mmax}{\Omega+i\eta\mmax}}}}a_{(m-1)0}.
\end{equation}
A similar relation is obtained for $n=1$. Since the recursion relation in \eqr{\ref{eq:linKinEqProj}} does not couple Laguerre moments together, one need only solve the system for $a_{00}$, $a_{10}$, $a_{20}$ and $a_{01}$. The coefficients $a_{30}$ and $a_{11}$ can be written in terms of continued fractions like \eqr{\ref{eq:cFrac1}}, and one obtains a linear problem with the determinant of the mass matrix yielding the dispersion relation~\cite{Anderson2007}
\begin{equation} \label{eq:ionAcousticDR}
\frac{T_{i0}}{T_{e0}} = \frac{8\eta^2+9F_1F_2}{8\eta^2\left(\Omega^2-1\right)+9\left[\left(\Omega^2-1\right)F_1-2\Omega\right]F_2},
\end{equation}
where the functions $F_1(\Omega,\eta)$ and $F_2(\Omega,\eta)$ are
\begin{equation}
\begin{aligned}
F_1\left(\Omega,\eta\right) &= \Omega+\cfrac{4}{3}i\eta-\cfrac{3}{\Omega+3i\eta-\cfrac{4}{\Omega+4i\eta-\cfrac{5}{\Omega+5i\eta-\dots \cfrac{\mmax}{\Omega+i\eta\mmax}}}}, \\
F_2\left(\Omega,\eta\right) &= \Omega+\cfrac{2}{3}i\eta-\cfrac{1}{\Omega+3i\eta-\cfrac{2}{\Omega+4i\eta-\cfrac{3}{\Omega+5i\eta-\dotsc\frac{\mmax}{\Omega+i\eta\left(\mmax+2\right)}}}}.
\end{aligned}
\end{equation}

\begin{figure}
  \centering
  \includegraphics[width=0.49\textwidth]{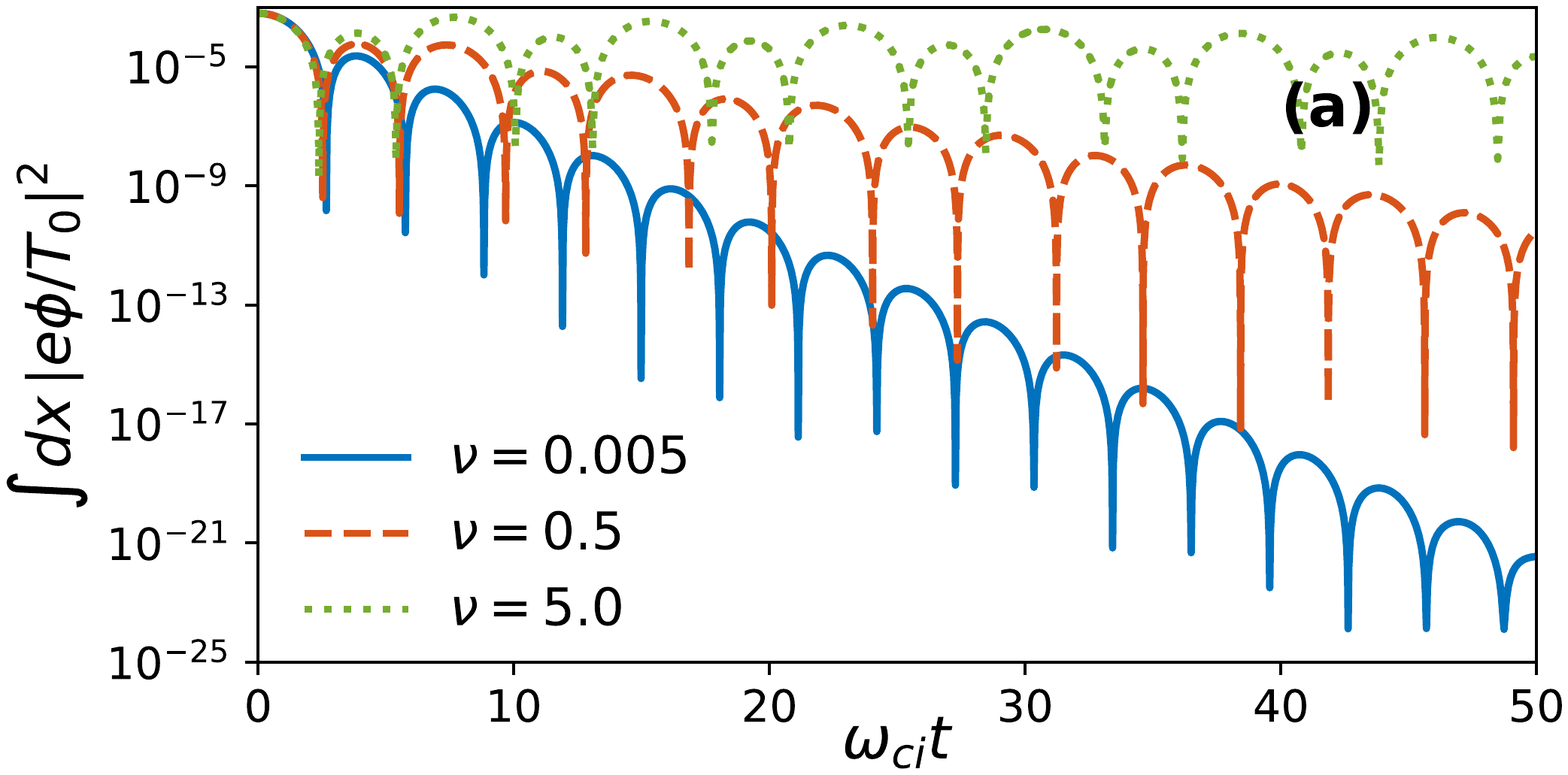}
  \includegraphics[width=0.49\textwidth]{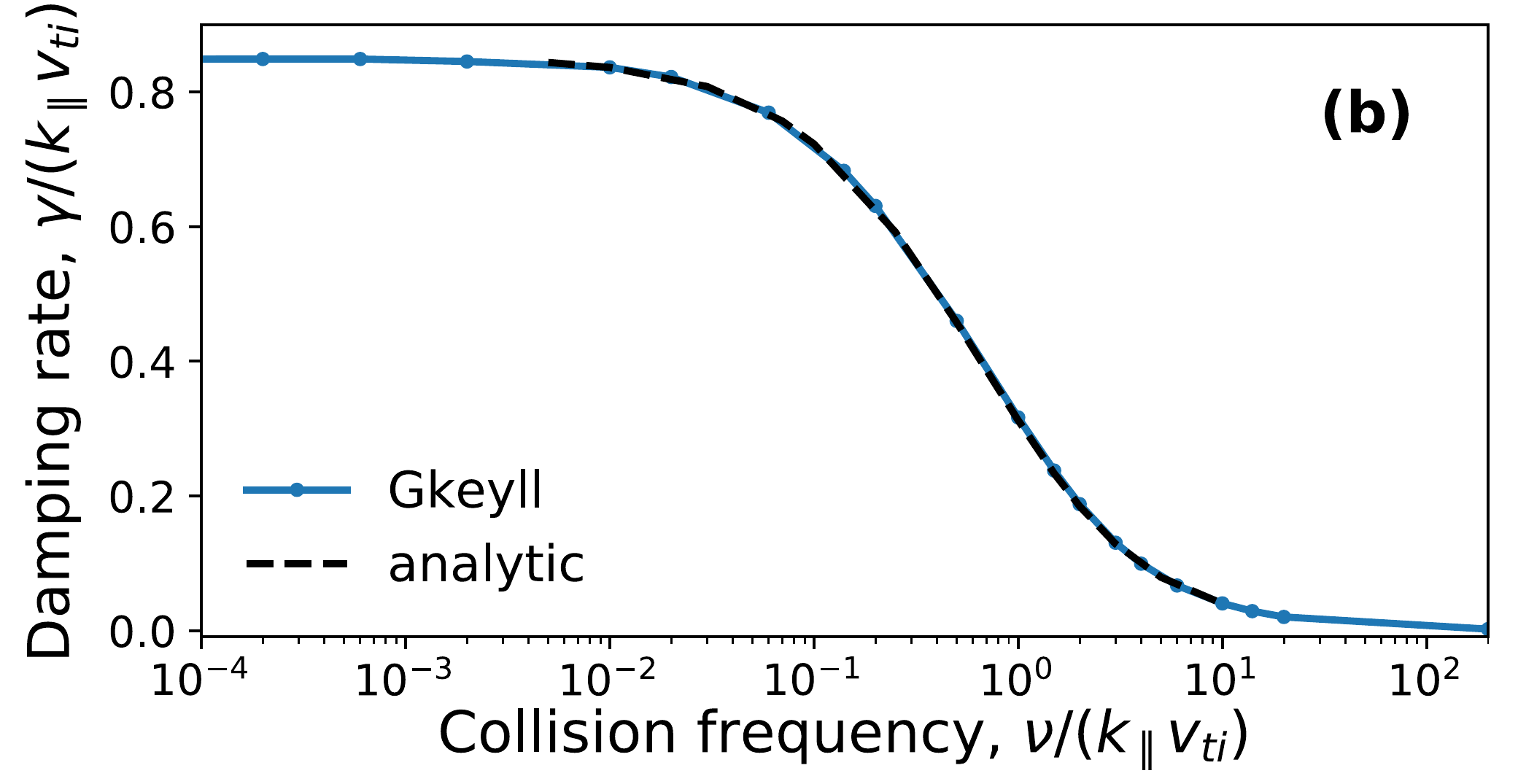}
  \caption[1X2V Landau damping of ion acoustic waves.]{(a) Field energy time trace and (b) damping rates as a function of collisionality for the ion acoustic wave.}
  \label{fig:ionAcousticDamping}
\end{figure}

We set up an analogous scenario in \gkeyll~using adiabatic electrons, $\tau=1.0$, hydrogen mass ratio and perturbed the initial state using a wave mode with $\kpar\rhoi=0.5$. These simulations were done on a domain $[-\pi/\kpar,\pi/\kpar]\times[-6\vti,6\vti]\times[0,m_i(5\vti)^2/(2B)]$ discretized with $64\times128\times16$ cells. This resolution and the time step stability constraints are probably conservative and were chosen to guarantee these results were well converged. In \figra{\ref{fig:ionAcousticDamping}}{a} the decay of the wave is displayed by the decrease in electrostatic energy over time, from which one can measure both the real part and the imaginary part of the wave frequency. The energy trace of three different collisionalities show that as ion-ion collisions alone become more frequent, the damping mechanism is progressively eroded. We scanned the entire collisional range and compared our results to the roots of the dispersion relation in \eqr{\ref{eq:ionAcousticDR}}. \Figra{\ref{fig:ionAcousticDamping}}{b} shows excellent agreement between theory and our implementation in \gkeyll. Consistent with intuition, when the mean-free-path becomes comparable to the wavelength, $\nu/(\kpar\vti)\sim1$, fewer particles will be able to resonate with the wave before experiencing collisional scattering, thus considerably reducing damping. In the high-frequency limit, collisions maintain the plasma in a local thermodynamic equilibrium so the plasma behaves like an ideal gas that has undamped compressional oscillations. This test also confirms an earlier partial agreement between theory and simulation of Landau damping of electron (Langmuir) waves due to a disparity in the dimensionality of the two~\cite{Hakim2019}.

\subsection{Helical open-field-line plasma turbulence} \label{sec:5dturbulence}

We now present a benchmark test of the full 5D (long-wavelength) gyrokinetic system, given by \eqsr{\ref{eq:gk-boltzmann}}{\ref{eq:5-poisson}}. 
With the moment-conserving \mFPO, we simulated plasma turbulence on helical, open field lines, using a nonorthogonal field-line-following coordinate system as in~\cite{shi2019full,bernard2019}. In this coordinate system, $z$ is parallel to magnetic field lines, $x$ is the radial coordinate, and $y$ is the ``bi-normal" coordinate. To ensure numerical stability, we used \eqr{\ref{eq:omegaH-max}} to determine that \cfl\ $\approx 0.28$ was necessary to prevent the electrostatic shear Alf\'en mode from becoming unstable. We set \cfl~$=0.2$ as a conservative estimate. We used the same physical parameters as in~\cite{bernard2019} to simulate the Texas Helimak simple magnetized torus experiment and make direct comparison with those results. We calculated the ion and electron collision frequencies from~\cite{huba2009nrl} using background densities ($n_0$) and temperatures ($T_{s0}$) that are constant in space and time. It is important to note that simulations in~\cite{bernard2019} included collision frequencies with spatially- and time-varying densities and temperatures, as well as electron-ion collisions (but no ion-electron collisions). Those simulations were also carried out with an earlier, nodal DG scheme employing different algorithms than those presented here while still remaining conservative by correcting for the errors. We denote this by $\nu_{ss'}(\mathbf{x},t)$ to differentiate it from simulations presented here with the moment-conserving \mFPO, using the constant collision frequency $\nu_s$ and neglecting multi-species collisions. We also present results from a simulation with a reduced collision frequency, $0.1\nu_s$. All simulations were run to 16 ms. Calculated equilibrium profiles were averaged in time from 10 to 16 ms and in the bi-normal direction $y$.

\begin{figure}
    \centering
    \includegraphics[width=\textwidth]{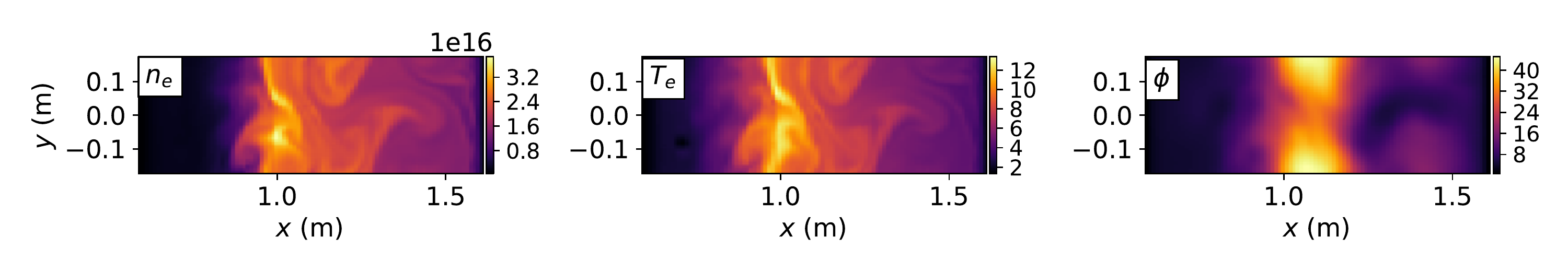}
    \caption{Snapshots of electron density (left), electron temperature (middle), and plasma potential (right) in the $xy$-plane, from simulations of plasma turbulence on helical, open field lines in 5D with the moment-conserving \mFPO.}
    \label{fig:ntphi-g2}
\end{figure}

\begin{figure}
    \centering
    \includegraphics[width=\textwidth]{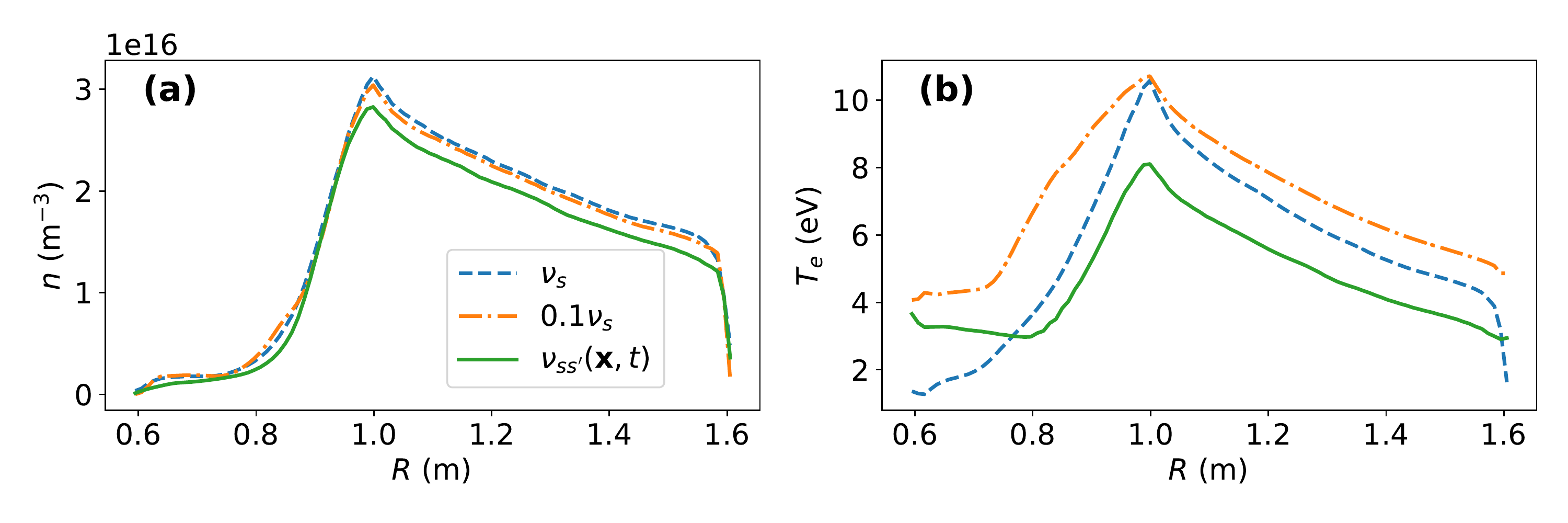}
    \caption{Comparison of (a) electron density and (b) electron temperature equilibrium profiles from simulations with different constant collision frequencies ($\nu_s,0.1\nu_s$) to that with time- and spatially-varying collision frequencies and multi-species collisions ($\nu_{ss'}(\v{x},t)$).}
    \label{fig:nT-g2}
\end{figure}

\Figr{\ref{fig:ntphi-g2}} shows snapshots of electron density, electron temperature, and plasma potential in the nonorthogonal field-line-following coordinate system at 10 ms. Turbulent structures and density levels are very similar to those presented in \cite{bernard2019}, though electron temperature and plasma potential values are slightly greater. Electron density profiles are compared in~\figra{\ref{fig:nT-g2}}{a}, with all three profiles being very similar. More differences are visible in~\figra{\ref{fig:nT-g2}}{b}, which compares the electron temperature profiles. Compared to the simulation with constant like-species collisionality only (dotted blue line in~\figra{\ref{fig:nT-g2}}{b}), including electron-ion collisions and spatially varying collisionality (solid green line~\figra{\ref{fig:nT-g2}}{b}) reduced the electron temperature. Given the inverse dependence of the interchange linear growth rate on the electron-ion collisionality~\cite{Ricci2009} one may consider the possibility of cross-field transport increasing as $\nu_{ei}$ decreases; were this effect to be significant parallel transport would be less competitive against perpendicular fluxes and would not carry out heat through the sheath as efficiently, leading to a temperature increase across the plasma. However the linear analysis suggests that the interchange growth rate is only weakly dependent on $\nu_{ei}$~\cite{Ricci2009}. Instead, a contributing factor to the increase of $T_e$ when collisions strengthen is that due to pitch-angle scattering more electrons are carried to higher $\vpar$, where they are lost through the sheath. This effect can increase the heat loss rate, lowering the temperature of the remaining electrons. Such mechanism would also apply to the increase in $T_e$ seen in comparing the simulation using constant like-species collisions (dotted blue line in~\figra{\ref{fig:nT-g2}}{b}) with a similar simulation which used a reduced collisionality (orange dash-dot line in~\figra{\ref{fig:nT-g2}}{b}). 


\iffalse
% MF: This paragraph is rewritten immediately below.
Plasma potential profiles are compared in~\figra{\ref{fig:phi-dn-g2}}{a}. All three $\phi(R)$ profiles are relatively similar, yet it seems that compared to the $0.1\nu_s$ run increasing the collisionality can lead to a magnification of the $E{\times}B$ shear. A stronger velocity shear contributes to the weakening of the linear growth rates and cross-field transport. It is also interesting that given the electron temperature profiles in \figra{\ref{fig:ntphi-g2}}{b} and assuming an adiabatic electron response, $e \phi \sim \Lambda T_e$, one might expect the potential profile from the simulation with lower collisionality to be higher than the others, which is not the case. It is possible that the lower collisionality allows for a non-adiabatic response of the plasma potential. Density fluctuation levels in~\figra{\ref{fig:phi-dn-g2}}{b} are reduced for the lower collision frequency case as compared with the other simulations. In general, the overall comparison for these quantities indicates that the moment-conserving \mFPO~reproduces previous simulations reasonably well. This will likely improve by including features such as spatially-varying collision frequencies and multi-species collisions.
\fi

\begin{figure}
    \centering
    \includegraphics[width=\textwidth]{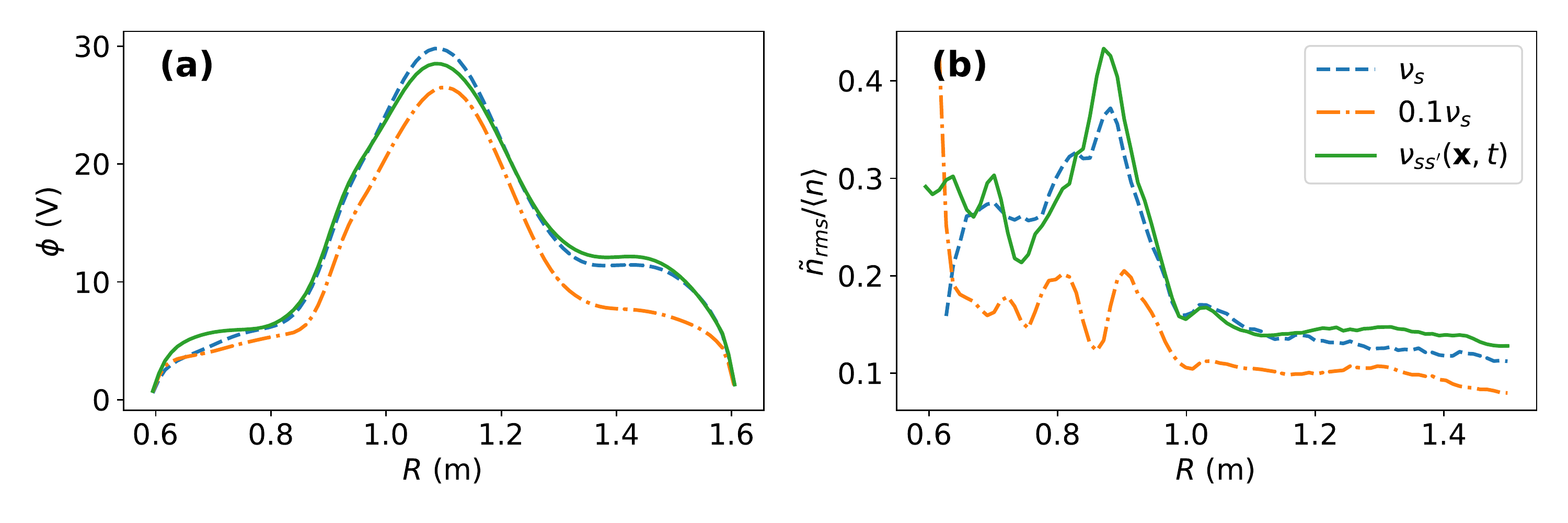}
    \caption{Comparison of (a) plasma potential and (b) density fluctuation profiles from simulations with different constant collision frequencies ($\nu_s,0.1\nu_s$) to that with time- and spatially-varying collision frequencies and multi-species collisions. A non-adiabatic electron response might explain the slight decrease in the plasma potential in the lower collisionality case. Turbulence levels in (b) are also reduced for the lower collisionality case.}
    \label{fig:phi-dn-g2}
\end{figure}

Plasma potential profiles are compared in~\figra{\ref{fig:phi-dn-g2}}{a}. All three $\phi(R)$ profiles are relatively similar, and any collisionality-induced changes to the $E{\times}B$ profile do not appear significant enough to indicate that shear stabilization would play a major role in the changes to the simulated profiles or the turbulence. It is however interesting that the constant like-species collisionality simulation with higher $T_e$ (orange dash-dot line in~\figra{\ref{fig:nT-g2}}{b}) is actually the one with a lower potential, contrary to what we would expect from an adiabatic electron response $e \phi \sim \Lambda T_e$. One possibility is possible that the lower collisionality allows for an increasingly non-adiabatic response of the plasma potential. It is also possible that despite the increase in $T_e$ there is a stronger decrease in $\Lambda$: at low collisionality fewer electrons scatter above the sheath potential, so the sheath potential has to drop to allow more electrons to escape in order to match the ion flux into the sheath. Lastly, we highlight that density fluctuation levels are reduced for the lower collision frequency case as compared with the other simulations (\figra{\ref{fig:phi-dn-g2}}{b}).

A more in depth analysis of the physics of these simulations is possible but beyond the scope of this manuscript. In general the intention here is to demonstrate that the moment-conserving \mFPO~presented in this work has been successfully incorporated into more complex 5D simulations, and that despite being limited to like-species collisions it produces results with reasonable agreement with previous simulations~\cite{bernard2019}. This agreement will likely improve by including features such as spatially-varying collision frequencies and multi-species collisions.

%% file: conclusions.tex
\section{Discussion and summary} \label{sec:conclusion}

We have presented a gyroaveraged Lenard-Bernstein-Dougherty collision operator (\mFPO), including a novel formulation of the discrete discontinuous Galerkin form and its implementation in \gkeyll. Building upon~\cite{Hakim2019}, we use the concept of weak equality to formulate a recovery DG (RDG) algorithm for the diffusion term of the \mFPO. It also provides a rigorous means to compute the primitive moments, $\upar$ and $\vt$. If such calculations are carried out using point-wise or cell average-based operations, significant errors ensue, causing non-conservation and instability. This concept guarantees that our discrete operator retains conservation properties and leads to an energy-conserving scheme even in the case of piecewise linear basis functions, provided that we carefully consider quadratic quantities projected onto the $p=1$ basis. Weak equality is also crucial in the definition of spectral transforms of DG data.

The continuous \mFPO\ is self-adjoint and satisfies the $H$-theorem but we have not yet proven that the discrete operator retains such properties. This is challenging because the present discrete operator does not guarantee $f>0$, though we have already implemented a positivity-preserving drag term (not presented here). Self-adjointness enhances the efficacy of some approaches to accelerate the time integration (e.g. super time-stepping~\cite{Meyer2014}), which we eventually wish to implement in order to more efficiently model highly collisional plasmas. Guaranteeing positivity, self-adjointness and non-decreasing entropy in the DG discretization scheme is the objective of on-going work.

We analyzed the stability conditions for DG advection and diffusion problems, and used this to establish the time step stability criterion for the \mFPO. Satisfying these conditions helps to avoid some issues associated with negative values of the distribution function, since $f > 0$ is not currently guaranteed in our scheme. For the SSP-RK3 time integration in \gkeyll, we presented a conservative estimate of the largest, stable time step.

Relaxation tests of the pure \mFPO~demonstrated the exact numerical conservation properties of our scheme. These systems evolved to a maximum entropy solution which, as shown through a Hermite spectral lens, is subtly different from a Maxwellian projected onto the DG basis.  This makes the Hermite analysis of collisional DG data more complicated for the larger Hermite moments, since the high-$m$ noise associated with the evolution of the zeroth-order Hermite moment causes higher moments to deviate from the analytic solution. However, lower moments of the \mFPO~evolve according to analytic theory, and the evolution of higher moments converges with resolution.

We performed tests of collisional Landau damping of ion acoustic waves. Using a Hermite-Laguerre basis, we obtained a dispersion relation whose least-damped roots agreed well with simulation results. We presented a more complicated test of 5D turbulence on open, helical field lines in the Texas Helimak device, which agreed well with previous simulations, even without the additional physics of multi-species collisions and spatially-varying collision frequencies. Multi-species collisions give rise to slightly different (discrete) conservation laws and requirements for the calculation of the velocities and thermal speeds in the cross-collision terms. More accurate gyrokinetic simulations of laboratory and astrophysical plasmas will include spatially-varying, and even velocity-dependent, collisionalities. These features are currently being developed and tested within the \gkeyll\ code.

%% file: appendixA.tex
\section{Accessing and running \gkeyll} \label{sec:appendixGkeyll}

The \gkeyll~code (in binary and source format) and the input files to reproduce results presented here are available for download. \gkeyll~installation instructions can be found on the \gkeyll~website (\url{http://gkeyll.readthedocs.io}). The code can be installed on Unix-like operating systems (including Mac OS and Windows using the Windows Subsystem for Linux) either by installing the pre-built binaries using the conda package manager or building the code via sources. The input files for simulations presented here can be found at \url{https://github.com/ammarhakim/gkyl-paper-inp/tree/master/GkLBO}.

%The 5D Helimak simulations require a distributed memory, multi-node cluster to run on practical times, but every other simulation presented here can be run on a laptop.